\documentclass[10pt,twocolumn]{article}

\usepackage[utf8]{inputenc}
\usepackage[T1]{fontenc}
\usepackage{textcomp}
\DeclareUnicodeCharacter{2212}{\textminus}
\usepackage{newunicodechar}

\usepackage[margin=0.85in,columnsep=20pt]{geometry}
\usepackage{amsmath,amssymb}
\usepackage{graphicx}
\graphicspath{{Figures/}{Fig/}{./}}
\usepackage{booktabs}
\usepackage{threeparttable}
\usepackage{caption}
\usepackage{setspace}
\usepackage{authblk}
\usepackage[protrusion=true,expansion=false]{microtype}
\usepackage[hidelinks]{hyperref}
\usepackage{csquotes}

\usepackage[style=apa, backend=biber, natbib=true, uniquename=false]{biblatex}
\DeclareLanguageMapping{english}{english-apa}
\let\cite\citep   

\newcommand{\alttext}[1]{}
\newcommand{\ORCID}[1]{}   
\providecommand{\keywords}[1]{\vspace{0.5em}\noindent\textbf{Keywords:} #1\par}

\usepackage{array}
\usepackage{enumitem}
\usepackage{float}
\usepackage{ragged2e}
\usepackage{xcolor}
\usepackage{longtable}
\usepackage{rotating}
\usepackage[most]{tcolorbox}
\usepackage{fancyhdr}

\fancypagestyle{sipage}{%
  \fancyhf{}%
  \fancyhead[L]{\footnotesize\itshape Supplementary Materials}%
  \fancyhead[R]{\footnotesize\itshape Daryani, Bogen, \& Daepp}%
  \fancyfoot[C]{\footnotesize\thepage}%
}

\newcolumntype{L}[1]{>{\RaggedRight\arraybackslash}p{#1}}
\newcolumntype{C}[1]{>{\Centering\arraybackslash}p{#1}}
\newcolumntype{R}[1]{>{\RaggedLeft\arraybackslash}p{#1}}
\newcommand{\dg}{\ensuremath{^{\circ}}}
\newcommand{\mn}{\ensuremath{-}}
\newsavebox{\tabbox}
\newcommand{\tabnote}[1]{%
  \vspace{0.35em}%
  {\footnotesize\RaggedRight\textit{Note.} #1\par}}
\newtcolorbox{promptbox}[1][Prompt]{%
  breakable, enhanced, sharp corners=downhill, boxrule=0.4pt,
  colback=black!3, colframe=black!45,
  coltitle=white, colbacktitle=black!45,
  fonttitle=\footnotesize\bfseries, fontupper=\footnotesize,
  title={#1}, left=6pt, right=6pt, top=4pt, bottom=4pt,
  before skip=8pt, after skip=8pt}

\title{\bfseries Accurate in space, unreliable in time: how LLMs represent national cultural change}

\author[1]{Yalda Daryani}
\author[2]{Miranda Bogen}
\author[2]{Madeleine I. G. Daepp}
\affil[1]{Department of Psychology, University of Southern California}
\affil[2]{Center for Democracy and Technology}
\affil[ ]{\vspace{0.3em}\small Correspondence: \href{mailto:daryani@usc.edu}{daryani@usc.edu}}

\date{}

\begin{document}

\makeatletter
\twocolumn[
\begin{@twocolumnfalse}
\maketitle

\begin{abstract}
\noindent Assessments of cultural alignment have become an important part of the development and improvement of large language models (LLMs). However, the majority of the evaluations treat culture as a single snapshot, investigating only whether a model represents a society accurately at the current time. Research in cultural psychology shows that cultural values change at different rates and directions over time. Therefore, a ``culturally aware'' model should capture not only where a culture is today but also how it has changed over time. We examine this missing dimension of cultural awareness using more than two decades of the World Values Survey data. We compare the cultural trajectories of 40 countries with the trajectories produced by four state-of-the-art (SOTA) LLMs on the Inglehart–Welzel cultural map. Our findings show that while models generally place countries close to their most recent surveyed positions, these representations tend to lag several years behind that position. They also capture only part of the magnitude of the observed change, introduce movement where little occurred, and rarely reproduce reversals in countries’ trajectories. These findings point to temporal flattening and suggest that snapshot accuracy can give an incomplete picture of cultural awareness in LLMs and have implications for model evaluation, representational harms, and the governance of culturally aware AI systems.
\end{abstract}

\vspace{1.5\baselineskip}
\end{@twocolumnfalse}
]
\makeatother

\section{Introduction}

Half a century of social and cultural psychology research has established a foundational premise that societies do not sit still \citep{jackson2024worldwide, kurdi2025international}. Modernization has broadly pulled many societies from traditional toward secular-rational values and from survival toward self-expression values \citep{inglehart2000modernization, inglehart2005modernization, welzel2013freedom}. These broad trends unfold differently across societies, however. Individualism, for instance, has risen across nearly every society studied, yet at sharply different speeds \citep{santos2017global}. Movement across values can also reverse. Several Soviet successor states shifted toward survival and traditional values after the economic collapse of the early 1990s \citep{inglehart2000modernization}, and, in a number of advanced democracies, rising self-expression values provoked a traditionalist counter-movement among cohorts left behind by that shift \citep{norris2019cultural}. Because societies have moved at such different rates and in such different directions, the surveyed record of the past four decades shows values diverging globally rather than converging on a common destination \citep{jackson2024worldwide}. Therefore, a society is better described by a trajectory---a direction, a rate, and an occasional reversal---than by any single snapshot in measured values. 

Cultural values are now studied not only in populations but also in the LLMs that people across the world are increasingly using for both personal and professional tasks. In order to represent cultures accurately, LLMs must take into account the fact that cultures change. A given model may correctly represent the values of a society at a given time, but these values will not necessarily remain the same in the future. To determine whether an earlier representation of a given society is still relevant today and how this society might evolve in the future, it is necessary to have information on how the society has evolved in the past. Thus, in addition to the question of whether LLMs know enough about the cultures of all the different societies, there is also the question of whether these models represent cultures as in motion over time. The extent to which LLMs accurately represent cultures is an important question both for everyday people and for science. LLMs are used daily by hundreds of millions of people worldwide \citep{microsoft2026aidiffusion}, mediating how those people retrieve and interpret information \citep{zhu2025large} and potentially reshaping language and even cultural norms \cite{daryani2026homogenizing}. In social scientific research, moreover, LLMs have been advanced as low-cost ``silicon samples'' that can stand in for human populations \citep{argyle2023out}. Whose values a model reflects, then, is not only a scientific question about the validity of using LLMs to represent the human population but also a consequential governance question about representation, cultural erasure, and hegemony \citep{shelby2023sociotechnical, mollema2025taxonomy, de2024become}.

Early studies found strong and consistent evidence that model outputs frequently resemble the opinions of the United States and of English-speaking and Protestant-European populations \citep{durmus2024towards, tao2024cultural, santurkar2023whose}. This bias is observed across different benchmarks and measures. On standard psychological measures, model responses fall outside the range of global data \citep{pataranutaporn2025simulating} and track most closely with Western, Educated, Industrialized, Rich, and Democratic (WEIRD) populations \citep{atari2023which, cao2023assessing}. Changing the language in which LLMs are prompted does not consistently reduce this cultural skew \citep{mora2025art, bulte2026llms}, though more recent studies show that naming a country explicitly can shift model responses toward the values of that society, with cultural alignment being stronger for countries whose languages are better represented in a model's training data \citep{tao2024cultural, alkhamissi2024investigating}. 

Although there are accumulating studies examining how models represent cultures at a fixed point in time, we know of no study that asks whether models capture how a society has changed. Cultural alignment audits compare a model against a single recent snapshot, a design that treats a nation's culture as a fixed point rather than a trajectory \citep{masoud2025cultural, yuan2024comparative}. One study \citep{liu2026alignment} does recognize the importance of change, scoring models against WVS waves 5 through 7 and reporting that alignment is highest at the most recent wave for the five countries with which the models already align well. This is early evidence that models carry a recent picture of a society, but it leaves open whether and to what extent they capture change when prompted appropriately. 

Cultural alignment studies often draw on how culture is conceptualized in social sciences while neglecting an important part of that understanding, which defines culture as a process that is transmitted and revised over time, rather than being a fixed property of a nation \citep{kashima2019psychology, hong2000multicultural, oyserman2011culture}. Treating a society as unchanging can also foster an essentialist view of social groups where characteristics observed at one point in time are taken to be stable features of the group \citep{haslam2000essentialist}. This shortcoming could be consequential as cultural evaluations are incorporated into model documentations and assessments \citep{solaiman2025evaluating}, and cultural prompting is recommended as a strategy for improving alignment on these evaluations \citep{tao2024cultural}. Thus, cultural alignment evaluations require asking both whether models distinguish among societies and whether they capture how those societies change over time. 

In this paper, we distinguish between \emph{snapshot alignment} and \emph{change alignment}. Snapshot alignment asks whether an LLM locates a society correctly at a recent point in time by comparing its elicited values with that society's most recent results in the World Values Survey (WVS), the largest and longest-running empirical study of cultural values across countries. Change alignment captures whether and to what extent an LLM can capture how a society's values have changed over time, in terms of direction, rate of change, and reversals, again compared to the WVS. WVS items are commonly reduced to the two dimensions of the Inglehart–Welzel cultural map (traditional vs. secular-rational values and survival vs. self-expression values). The first dimension captures the differences in the importance placed on religion, obedience to authority, and traditional family norms, and the second dimension contrasts a focus on economic and physical security with a focus on tolerance, gender equality, participation, and quality of life. We test four models---Claude Opus 4.8, GPT-5.5, Gemini 3.6 Flash, and Qwen3.6-Plus---on ten core WVS indicators for 40 countries across Waves 4--7, using coordinates estimated directly from the survey microdata as our empirical benchmark. 

We begin by asking whether models place societies close to their most recent surveyed positions (snapshot alignment). We then look at time, asking which survey wave a model's representation of a country most closely resembles when no year is given in the prompt and how far behind the most recent survey position it is (identifying its ``implicit anchor'' year). We next evaluate the extent to which explicitly naming the year moves a model's representation closer to the country's surveyed position in that year. We examine whether models recover the direction and rate of observed cultural change and, importantly, whether their movement corresponds to when that change actually occurred rather than simply following the broad modernization trend shared across many societies. Finally, for societies whose surveyed trajectories reverse, we ask whether models reproduce that reversal or continue moving the country in its earlier direction. 

\section{Materials and Methods}
\subsection{Study design}
We compared LLM representations of national culture against repeated survey measurements in order to assess the extent to which LLMs captured cultural values for a given country at a single point in time (snapshot alignment) versus over time (change alignment). We drew our empirical benchmark data from the World Values Survey \citep{inglehart2022wvs}, eliciting LLM representations for the same items. We then projected survey and LLM estimates onto the same coordinate system so that the two can be directly compared for every country and wave. 

\subsection{Data}
\subsubsection{Survey data and country sample}

We used WVS trend data for Waves 4 through 7, covering fieldwork years from approximately 1999 to 2023. For estimating the trajectory of change parameters (i.e., direction, rate and reversal), we required at least three time points. Therefore, we only included countries that had been surveyed at least three times during this period and were also part of Wave 7, which served as our reference point for snapshot alignment. We did not require all countries to have been surveyed in every single wave, as this would have biased our sample towards countries with more resources and better coverage. Moreover, we did not exclude countries that were missing data for some WVS items in some waves. This was because the missing items were concentrated on politically and morally sensitive questions in precisely the authoritarian, religious, and post-conflict societies whose trajectories are of most interest. Therefore, including the item completeness requirement would have selected the sample on a variable correlated with the estimated outcome. We accounted for uneven item coverage by keeping the indicators consistent within each country, scoring model outputs on a country-by-country basis, and checking the results with sensitivity analyses. 

Applying the screen to the microdata yields 40 countries and 134 country-waves, with 14 countries observed at four waves and 26 at three. Countries were selected based on their WVS coverage, without considering the shape of their trajectories or their representation in the anglophone information environment. We classified their trajectories only after the sample had been selected, using the procedure described below. The full country list, wave coverage, and fieldwork years are provided in the SI. 

\subsubsection{Models and elicitation}
We elicited cultural representations from four large language models: Claude Opus 4.8, GPT-5.5, Gemini 3.6 Flash, and Qwen3.6-Plus.

We used the ten key indicators from the WVS with their response scales to query our models. Eight of the indicators were ordinal single-scale items (for example, a 1–10 justifiability scale or a 1–4 national-pride scale). We also included the child-qualities card sort, where participants were shown a set of cards with different qualities that children can learn at home, and asked to select up to five of these cards. Four of these qualities (independence (A029), determination and perseverance (A039), religious faith (A040), and obedience (A042)) were used to create a map, and they were scored based on how often people mentioned them. Because respondents could select only five qualities, choosing one affected which of the other options they could choose. Therefore, we presented all the qualities together, like a card sort, instead of asking the models to rate each one separately. The full set included ten qualities in Waves 4–6 and eleven in Wave 7. Finally, we also included the WVS post-materialism index (Y002), in which respondents had to pick the two most important national goals from a list of four options. 

To elicit LLM-based representations on these indicators, we prompted a set of four models on these ten indicators. All queries used the same system prompt, held constant across models, conditions, and waves, which stated the task and instructed the model to answer as a probability distribution over the response options (full instrument in SI). Every item in the instrument was prompted to the models in three different anchoring conditions. C0 named neither country nor year (``a population of adults'') and was elicited once per model as a country-agnostic baseline. C1 named the country but not the year, eliciting the model's present-tense representation. C2 named both the country and the fieldwork year of a given wave and was elicited once for each retained country-wave. 

For each item, we asked models to give a probability distribution over the response options rather than repeatedly sampling individual answers and using those samples to construct a distribution. This approach, known as a verbalized distribution, has been shown to recover human opinion distributions more accurately than repeated sampling or token log-probabilities over the response options \citep{meister2025benchmarking}. Repeated sampling is also not well suited to our purpose because variation across model responses does not necessarily represent variation in the population. This approach also appears to be relatively insensitive to prompt wording. Using WVS items across countries, \citet{liu2026alignment} report correlations above 0.96 between distributions produced with their default and perturbed prompts. Given that a model does not state exactly the same distribution every time it is asked, we repeated each query k times and took the mean stated distribution as the estimate for each cell (model, country, condition, item, and wave), with k = 10.

Responses were treated as usable when they parsed to a well-formed distribution, either exactly or after renormalization within tolerance. Malformed, empty, or refused responses were discarded and re-elicited. For the ordered-pair post-materialism item, models occasionally returned distributions whose probabilities summed slightly above one. We renormalized these within a fixed tolerance and applied the identical normalization rule uniformly across all four models. Response validation and normalization rules are specified in full in SI. 

\subsection{Cultural map coordinates}
We projected each country-wave and each model's answers for that country-wave onto the two axes of the Inglehart–Welzel cultural map, traditional versus secular-rational (Dimension 1) and survival versus self-expression (Dimension 2). We followed the established Inglehart–Welzel map in selecting the ten core indicators, assigning their thirteen component variables to the two dimensions, and determining how each component was scored \citep{inglehart2000modernization, inglehart2005modernization}. Each dimension's components, including full item wording, WVS variable codes, response scales, and orientation indications, are provided in SI. We calculated each dimension using the equal-weight mean of its standardized components. Equal weighting is a rule that can be applied identically to survey respondents and model outputs, ensuring that both sides arrive at the map using the same calculations and that any discrepancy between them reflects the responses rather than the scoring.

WVS data and LLM responses differ only in how a single item value is obtained. On the survey side, an item value is a respondent's coded answer, with survey-defined missing codes set to missing. On the model side, each model's mean stated distribution for an item was converted into a value comparable to the corresponding WVS respondent mean. For the eight ordinal items, the item value was the expected value of the stated distribution (the sum of each response option multiplied by its probability), which is directly comparable to the population mean of that item. For the child-qualities card sort, we derived the four child-autonomy components as mentioned proportions, taking the model's stated selection probability for each of the four scored qualities. Because the deck differs across waves, we mapped each quality to its wave-specific position so that the correct qualities were scored. For the post-materialism item, we converted the model's distribution over ordered goal-pairs into the WVS post-materialism index value: each ordered pair was assigned to the index category corresponding to how many of its two goals were post-materialist, and the item value was the probability-weighted mean on the resulting 1–3 index.

From item values onward, the procedure is identical for both sides. Each component variable was z-scored using per-item means and standard deviations computed once over the pooled WVS respondent sample and sign-oriented so that higher values correspond to the secular-rational and self-expression poles. On the survey side, this standardization is applied at the respondent level and averaged to a country-wave national mean using the survey design weight (S017). Given that standardization is linear, this is equivalent to standardizing the weighted national mean, which is the form the model side takes. Each dimension score is the equal-weight mean of its sign-oriented standardized components. To keep a country's coordinate comparable across its own waves and to keep each model's coordinate comparable to the empirical benchmark for that country, we applied a constant within-country indicator base, where a component was retained for a country only if non-missing for more than half of respondents in all of that country's waves, and the same per-country base was applied when scoring model outputs. Nine countries had one or more components excluded, and their coordinates rest on fewer indicators and are less precisely estimated. We assessed the consequences by repeating the timing analysis on the countries with the complete battery (see Results). This yields, for each country-wave, one Dimension 1 and one Dimension 2 coordinate for the survey and one for each model in each condition, on identical axes.

\subsubsection{Trajectory classification}
We treated each country's ordered wave coordinates as a path in the two-dimensional cultural space and derived four quantities from it that allowed us to classify each country's trajectory shape and measure how far and in what direction models moved that country. \emph{Step length} is the distance between consecutive waves; \emph{net displacement} is the straight-line distance from a country's first surveyed position to its last; \emph{total path length} is the sum of the step lengths, which exceeds net displacement whenever a path bends; and the \emph{turn angle} at each interior wave is the angle between the incoming and outgoing steps. \emph{Straightness}, net displacement divided by total path length, equals 1 for a path that never changes direction and approaches 0 for one that returns toward its origin. This is the straightness index of trajectory analysis, also known as the net-to-gross displacement ratio \citep{benhamou2004reliably}. 

Sampling uncertainty was quantified by a nonparametric bootstrap. Within each country-wave, we resampled respondents with replacement, carrying the survey design weights, and recomputed that country's coordinates and the geometry of its entire path on each replicate. In this method, the surveyed values remain fixed across replicates, while the sample of respondents changes. Thus, the bootstrap distribution of straightness describes how precisely a country's path shape is determined given its sample sizes, and a country whose interval spans a wide range has a shape the survey data do not resolve.

Classification followed a two-stage rule, with a substantive verification stage applied afterward to the reversal cell. First, countries whose total path length per decade fell below 0.13 standardized units were classified as low-change/stable. The floor is expressed relative to the scale of the map. The average country sits 0.40 units from the global centroid, so a country at the floor covers roughly a third of that distance over a decade. We repeated the classification at floors of 0.10 and 0.16 and report the consequences in SI. Second, the remaining countries were classified from the bootstrap distribution of straightness: reversal if the interval's upper bound fell below 0.5, directional if its lower bound exceeded 0.7, and ambiguous otherwise. Using the interval rather than the point estimate helps avoid treating small shifts due to sampling noise as real changes in direction. Both cutoffs were fixed before the intervals were computed. Because they leave a gap between 0.5 and 0.7, a country is labeled only when its entire bootstrap interval falls below 0.5 or entirely above 0.7. A country whose interval falls anywhere in between or straddles either cutoff is left ambiguous rather than assigned to whichever class is nearer. Because these are hard thresholds on bootstrap percentiles, a country whose relevant bound sits near a cutoff could, in principle, flip class across Monte Carlo draws, so we repeated the classification under five bootstrap seeds and demoted to ambiguous any label that changed. At 2,000 replicates none did. Two near-boundary cases are recorded: Japan's straightness upper bound sits just below the reversal cutoff under every seed (0.490 to 0.498), and Germany's lower bound is just above the directional cutoff (0.706 to 0.712), with Germany's directional status additionally checked in a with/without sensitivity run.

Stage 2 yielded four stable countries, ten directional countries, seven candidate reversals, and nineteen ambiguous countries. The bootstrap resamples respondents within waves and therefore cannot distinguish genuine doubling-back from a turn that is merely orthogonal, since a right-angle turn also lowers straightness. We verified each candidate reversal on two further criteria. First, we required the turn at the bend to be obtuse, taking angles near or above 120° as back-tracking and angles near 90° as an orthogonal shift that only reads as low straightness. All seven candidates cleared this requirement, with turns from 129° to 169°. Second, we required shape concordance with an independent operationalization of the same dimensions: Welzel's secular values and emancipative values indices \citep{welzel2013freedom}, standardized on the same scale, from which we recomputed each country's path and its straightness. The question is whether a separately constructed measure also bends. Six candidates do, with straightness on the Welzel path ranging from 0.13 to 0.70 (Indonesia 0.13, India 0.21, Zimbabwe 0.23, Jordan 0.35, Japan 0.40, South Korea 0.70). Vietnam’s Welzel path runs essentially straight at 0.97, suggesting that the apparent reversal is specific to the thirteen-item composite, but its three-wave span does not allow for an additional comparison. We reclassified Vietnam as ambiguous, giving the twenty ambiguous countries reported in (Table~\ref{tab:trajectory_classes}). Japan is retained despite its near-boundary bootstrap margin, its 129° turn on the survival–self-expression axis corroborated by a concordant Welzel path.

\subsection{Analysis}
\subsubsection{Snapshot alignment}
We define snapshot alignment as the Euclidean distance in the two-dimensional map plane between a model's coordinate for a country and that country's most recent WVS coordinate, computed for the country-only condition (C1) and, as a year-anchored comparison, for C2. To detect directional bias, we decomposed the gap into its signed per-dimension components. We also tested whether models compress the map, representing countries as more similar to one another than they are. For each dimension, we computed the standard deviation of model coordinates across countries and divided it by the standard deviation of the corresponding empirical benchmark coordinates so that a ratio of 1 indicates a model separates countries as widely as the survey record does and a ratio below 1 indicates it places them closer together. 

\subsubsection{Change Alignment}
Change alignment comprises the direction of a model's movement, its rate, and its timing. For direction and rate, we computed the net displacement of each path (the start-to-end vector for the model and the empirical benchmark) and took the cosine between the two vectors and the ratio of their magnitudes. 

Because the WVS country paths all mostly drift in the same dominant direction to the secular and self-expression poles, a model that was only able to move every society in this same direction would get high net displacement scores for tracking this very same general direction, regardless of not actually tracking any of the countries in particular. Consequently, we computed alignment step by step: for each pair of consecutive waves, the cosine between the model's step and the true step, averaged over a country's steps. We measured this against a permutation null \citep{ernst2004permutation}, where for each step of the model, we paired it with a random survey year (i.e., a permutation that preserved the model’s general direction but otherwise was completely random with respect to the true country-specific paths and their corresponding years of change). We refer to this stepwise alignment subtracted by the corresponding permutation null mean as the ‘timing signal,' which we aggregate and then bootstrap across countries. A timing signal close to zero means that a model, even if it captures change for a country, in fact conveys no information as to when in that country’s particular path the change occurred. 

\subsubsection{Reversal reproduction}
For each of the six reversal countries, we calculated the turn angle at every interior wave. The angle was defined by the incoming and outgoing steps, with 0° indicating continuation in the same direction and 180° a complete reversal. We located the country's reversal at the wave with the largest turn in the WVS trajectory and then measured the model's turn at that same wave.

Turn angle alone, however, does not show how much movement occurred. Two very short steps can form a sharp angle even when the model's position changes very little. We therefore counted a reversal as reproduced only if the model turned at least 120° and the combined length of the two steps around the turn was at least half the distance observed in the WVS trajectory. Two of the six countries had only three survey waves, so their trajectories contained a single interior wave at which a turn could be measured.

\section{Results}
The analytic sample is 40 countries and 134 country-waves (Table~\ref{tab:trajectory_classes}). For each item, we asked the models to provide a probability distribution over the response options and averaged the distributions across ten independent elicitations. We used this approach rather than repeatedly sampling individual responses from the models following research showing that these verbalized distributions recover human opinion distributions more accurately than repeated sampling or token log-probabilities \cite{meister2025benchmarking, liu2026alignment}. They can also be obtained from closed APIs, unlike token log-probabilities, which allowed us to use the same elicitation procedure for all four models. 

We prompted each model under three conditions that differed in whether the country and year were specified. C0 named neither the country nor the year, C1 named the country only, and C2 named both the country and the fieldwork year of a given wave. We use C1 to examine how a model represents a country's values when no time period is specified. This condition is used for the snapshot and implicit-anchor analyses. For the snapshot analysis, we also report C2 to examine whether adding the year changes how closely the model matches the country's most recent surveyed position. The change analyses rely on C2 because comparing change over time requires the model and WVS data to refer to the same country and year. Finally, C0 provides a baseline for how models respond when no information about the country or year is given.

Before comparing model and country positions, we established the scale of the cultural map. Country coordinates fall roughly within ±0.70 standardized units on each axis, and the average country sits 0.40 units from the global centroid. These values provide a reference for the model-country distances reported below. 

Because the analyses that follow examine how countries move through the cultural map, we should first clarify what we mean by movement. A country's position along the two Inglehart–Welzel dimensions in a given survey wave can be compared to its position at the previous wave, enabling us to measure both the magnitude and direction of the change. We can then compare this observed movement with the movement produced by an LLM prompted to produce a response distribution for the same country and years. To evaluate how countries moved over time and how LLM representations compare to real-world WVS results, we calculated both the distance between coordinates on the Inglehart-Welzel map and the angle of change from one survey wave year to the next. An angle of 45° indicates some directional agreement, 90° means the two movements are perpendicular, and 180° means the model moves the country in the opposite direction. A schematic of both measures is provided in SI. 

We measured \emph{snapshot alignment} as the distance between a model's coordinate for a country and that country's most recent surveyed position. We also identified the \emph{implicit anchor}, or the survey wave whose coordinate was closest to the placement a model gave when asked about a country without specifying a year (C1). We measured \emph{change alignment} in three ways. First, for \emph{direction and timing}, we calculated the cosine between a model's step from one wave to the next and the country's true step over the same period and compared this with a permutation null that shuffled the model's steps across periods. Second, we measured \emph{rate} as the ratio of the distance a model moved a country to the distance that country actually moved. Third, for those countries with major \emph{reversals}, we measured the model's turn angle at the same wave where the country's sharpest documented turn occurred.

\begin{table*}[t]
\centering
\captionsetup{labelfont=bf, labelsep=newline, font=it,
              justification=raggedright, singlelinecheck=false}
\caption{Final Trajectory Classification of the 40 Countries}
\label{tab:trajectory_classes}
\begin{tabular}{p{2.5cm}cp{12.5cm}}
\toprule
Trajectory class & $n$ & Countries \\
\midrule
Reversal & 6 & Indonesia, India, Jordan, Japan, South Korea, Zimbabwe \\
Directional & 10 & Germany, Thailand, Taiwan, Canada, Kyrgyzstan, Romania, Turkey, Serbia, Brazil, Netherlands \\
Low-change/Stable & 4 & Russia, Colombia, Philippines, Cyprus \\
Ambiguous & 20 & Egypt, Iraq, Morocco, Malaysia, Hong Kong, Chile, Pakistan, Iran, China, Australia, Ukraine, United States, Argentina, Nigeria, Mexico, Vietnam, New Zealand, Peru, Uruguay, Singapore \\
\bottomrule
\end{tabular}
\end{table*}
 
\subsection{Snapshot alignment}

Given a query that named only the country (C1), all four models place countries close to their true recent position. Mean distance to a country's most recent WVS coordinate ranged from 0.153 to 0.238, against an average country-to-centroid distance of 0.40, indicating broadly accurate representation of each country's current values. These representations are, moreover, a significant improvement in accuracy compared to the naive baseline in which no country is named (C0). When no country was named, models sat 0.49 to 0.62 unit from the truth, and naming the country reduced that distance by 0.32 to 0.38 units (Claude 0.316 [0.256, 0.378], GPT 0.357 [0.284, 0.432], Qwen 0.378 [0.296, 0.463], Gemini 0.379 [0.307, 0.457]). Naming the country also relocated the answer by 0.39 to 0.55 units across the map (Claude 0.390, Qwen 0.453, GPT 0.486, Gemini 0.545), which exceeds the average country's 0.40-unit distance from the global centroid.

Accuracy varied more between countries than across models. Mean placement error ranged from 0.09 for Iran to 0.40 for Thailand, with four of the five countries that models placed least accurately located in East and Southeast Asia (Thailand, Taiwan, South Korea, and Japan). Comparing models, Gemini placed countries closest to their most recent surveyed position (0.153, 95\% CI [0.124, 0.182]), but its advantage over Claude was not statistically significant (difference 0.024 [−0.007, 0.055]). Claude (0.177 [0.146, 0.208]) and GPT (0.193 [0.160, 0.224]) were statistically indistinguishable from each other, while Gemini was significantly closer to the surveyed position than GPT (difference 0.040 [0.017, 0.065]). Qwen was the least accurate (0.238 [0.202, 0.278]), performing significantly worse than each of the other three under the paired bootstrap. 

Notably, the inclusion of the year in the model query had little effect on snapshot accuracy. Naming the country alone (C1) put models about as close to the recent snapshot as naming the country and year together (C2) (0.153 versus 0.151 for Gemini, 0.177 versus 0.182 for Claude, 0.193 versus 0.184 for GPT, and 0.238 versus 0.231 for Qwen). 

Model responses included distortions on both dimensions of the cultural map. On the self-expression axis, three of the four models placed countries significantly higher than the WVS records: Qwen by +0.118 ([0.067, 0.169]), GPT by +0.085 ([0.048, 0.124]), and Claude by +0.043 ([0.010, 0.076]). Gemini showed no significant displacement on this axis (0.000 [−0.029, 0.031]). The pattern differed on the tradition–secular axis. Claude (−0.049 [−0.098, −0.001]), Gemini (−0.047 [−0.093, −0.004]), and GPT (−0.046 [−0.092, −0.002]) placed countries significantly toward the traditional pole, while Qwen placed them toward the secular pole (+0.067 [0.018, 0.114]). Qwen differed significantly from each of the other three models on this axis (differences of 0.113 to 0.116, with all intervals excluding zero), while Claude, Gemini, and GPT did not differ significantly from one another. Thus, the self-expression bias appears in the same direction across models when it is present, while the tradition–secular bias differs by model. 

Finally, we asked whether models compress the cultural map by placing countries closer relative to their placement according to the WVS data. Comparing the spread of each model's coordinates with the empirical spread, we found evidence of compression in three of the eight model–axis combinations. Claude compressed the tradition–secular axis (ratio 0.65 [0.59, 0.73]) and, more marginally, the survival–self-expression axis (0.88 [0.80, 0.99]), while Qwen compressed the survival–self-expression axis (0.77 [0.67, 0.93]). No other interval differed significantly from 1. Overall, the models generally preserved the differences between countries observed in the WVS rather than pulling them toward a common center, although there was some evidence of compression for Claude and Qwen. 

\subsection{Implicit anchor}

\begin{figure}
    \centering
    \includegraphics[width=1\linewidth]{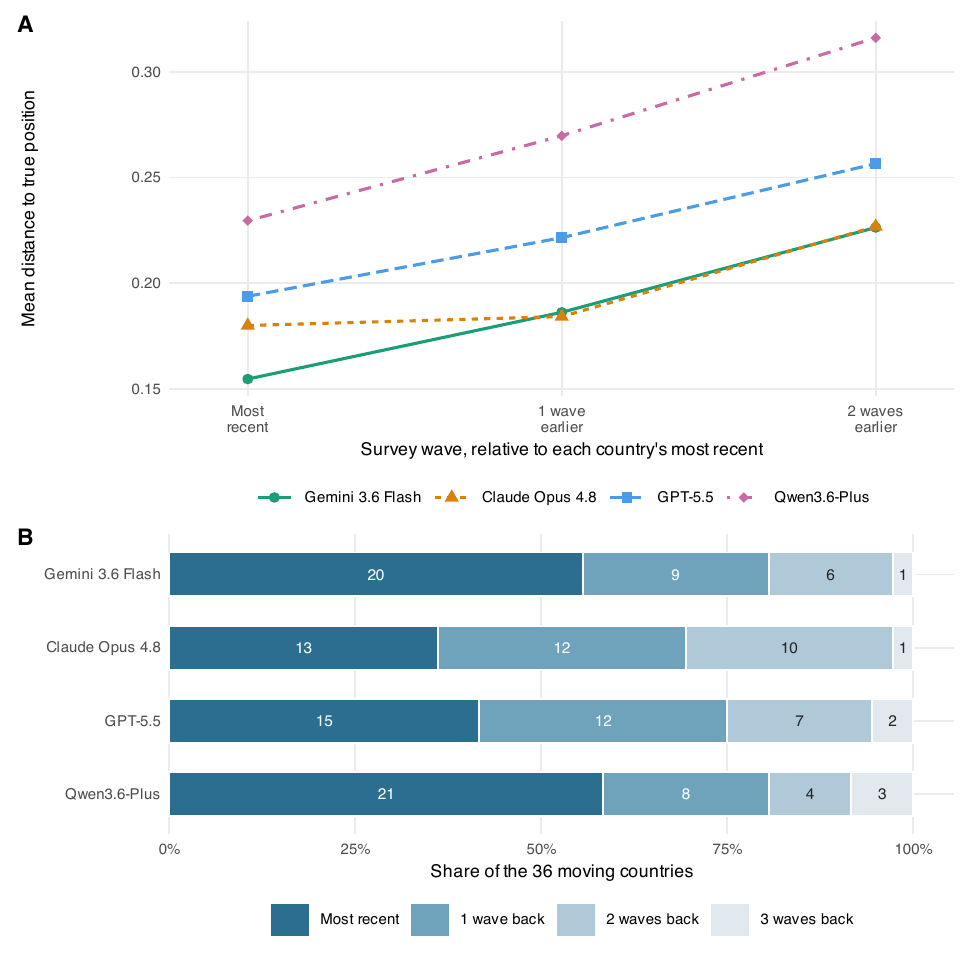}
    \caption{\textbf{The implicit temporal anchor.}
(A) Mean distance between each model's placement when no year is specified and the country's WVS coordinate at each survey wave, calculated across the 36 countries with measured movement. Survey waves are indexed relative to each country's most recent survey, and lower values indicate a closer match. (B) The closest survey wave to each model's placement for each of the 36 countries. Numbers show the number of countries assigned to each wave.}
    \label{fig1}
    \alttext{Two panels. Panel A is a line chart with one line per
    model showing mean distance to a country's true position rising steadily from the most recent survey wave to two waves earlier; all four models fit the most recent wave best, with Gemini 3.6 Flash closest throughout and Qwen3.6-Plus furthest. Panel B is a horizontal stacked bar chart, one bar per model, showing how many of the 36 countries anchor to each wave. The most recent wave accounts for 21 countries for Qwen3.6-Plus, 20 for Gemini 3.6 Flash, 15 for GPT-5.5, and 13 for Claude Opus 4.8, with the remainder spread across one, two, and three waves back.}
\end{figure}

We defined a model's ``implicit anchor'' for a country as the survey wave closest, in projected space, to the answer the model gives when asked about a country with no year given. We restricted this analysis to the 36 countries with measured movement. Averaged over those countries, every model's answer fit the most recent wave best, with fit degrading steadily as the waves got older (Fig.~\ref{fig1}A). However, assigning each country to its own closest wave, rather than averaging distances across countries, gave a more nuanced picture. The most recent wave was the best-fitting wave for only 36\% of countries for Claude and 42\% for GPT, compared with roughly 57\% for Gemini and Qwen (Fig.~\ref{fig1}B).

To further examine how far back each model's representation falls when it is closer to an earlier wave (versus the most recent wave), we measured the number of years between each country's most recent survey and the survey wave closest to the model's representation. Averaged across countries, Claude's anchor was 7.0 years behind the most recent survey (95\% CI [4.9, 9.3]), followed by GPT at 6.0 years [4.0, 8.1], Qwen at 4.7 years [2.6, 7.0], and Gemini at 4.5 years [2.6, 6.6]. Because WVS waves are unevenly spaced, we also measured the lag in number of survey waves rather than years. Using this metric, Claude anchored an average of 0.97 waves behind each country's most recent survey [0.69, 1.25], GPT 0.89 [0.61, 1.19], Qwen 0.69 [0.39, 1.03], and Gemini 0.67 [0.39, 0.94].

\subsection{Timing and rate}

Models generally shifted countries in the same broad direction as the WVS data, with median direction agreement ranging from 0.71 to 0.89 across models. The angle between the model's movement and the country's actual movement was less than 90° for 83\% to 92\% of countries. Using a stricter criterion of 45° reduced this proportion to 50–69\%.

\begin{figure}
    \centering
    \includegraphics[width=1\linewidth]{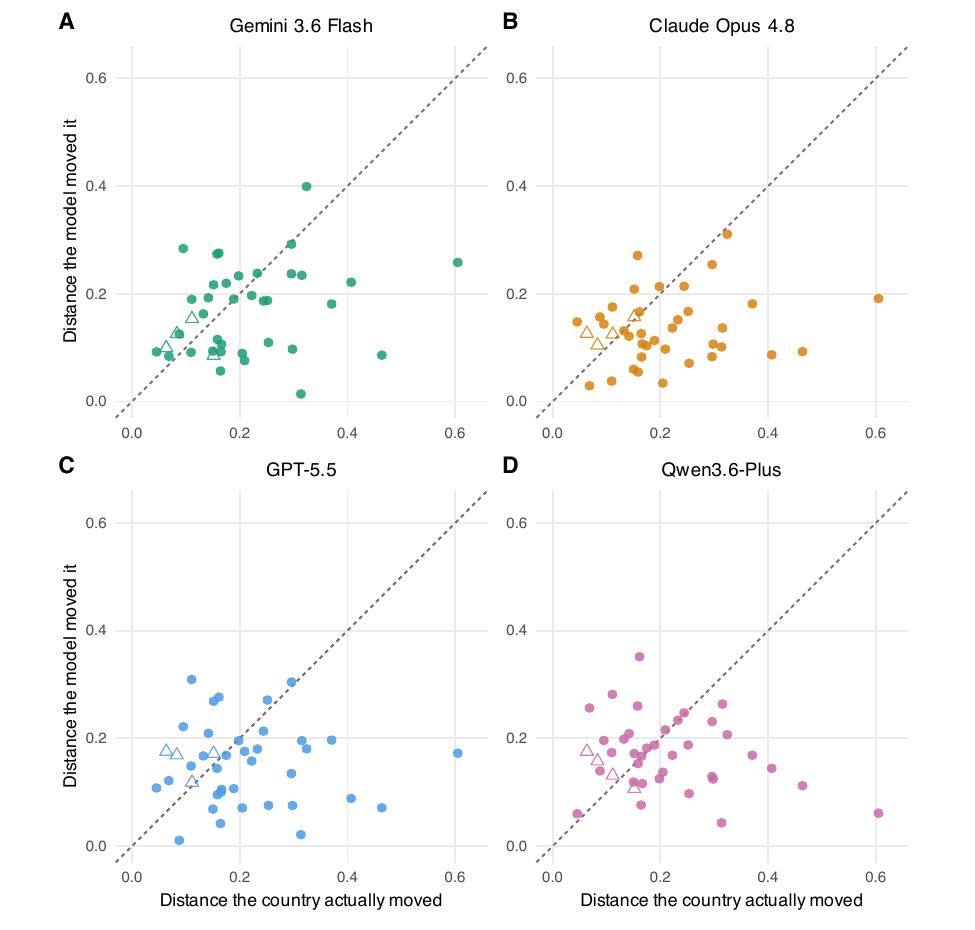}
    \caption{\textbf{Models under-travel where change happened and drift where it did not.} (A–D) Each point represents one country, with its observed net displacement on the cultural map between its first and last survey waves plotted against the corresponding net displacement in the model's representation. All 40 countries are shown: filled circles indicate the 36 countries with measured movement, and open triangles indicate the four countries classified as low-change. The dashed diagonal marks equal displacement in the model and WVS data; points below the line indicate that the model captured less movement than observed, while points above the line indicate more movement. Most countries fall below the line, while the four low-change countries fall above it.}
    \label{fig2}
    \alttext{Four scatter panels, one per model. In each, most
countries fall below a dashed diagonal line, indicating the model moved the
country a shorter distance on the cultural map than it actually moved. The four low-change countries, marked with open triangles, sit at the low end of the horizontal axis and mostly above the line, indicating movement the surveys do not record.}
\end{figure}

No model reflected the full amount of societies' changes. Figure~\ref{fig2} compares the observed and model-estimated net displacement for all 40 countries, separately for Gemini (A), Claude (B), GPT (C), and Qwen (D). For each country, we compared how far the model moved that country's position with how far the country's position changed in the WVS data. We expressed this as a ratio, where 1.0 means that the model captured the full magnitude of the observed change. Among the ten countries with a clear directional path, every model captured only part of the observed change, with median ratios of 0.45 for Claude, 0.56 for Gemini, 0.58 for GPT, and 0.68 for Qwen. This under-travel remained significant across all 36 movers for Claude (0.61 [0.43, 0.86]) and GPT (0.69 [0.55, 0.97]) but not for Gemini (0.82 [0.64, 1.18]) or Qwen (0.90 [0.70, 1.03]). 

Movement also appeared where little was observed in the survey data. The four low-change countries, shown as open triangles in Fig.~\ref{fig2}, moved a median of 0.097 units. Every model moved these countries farther, from 0.113 to 0.170 units. As Fig.~\ref{fig2} shows, the low-change countries fall above the equality line for all four models, indicating more model-estimated movement than observed movement. With only four countries in this group, however, we treat this pattern as descriptive rather than as a tested effect.

These measures compared a country's starting and ending positions, but nearly every country in our sample moved in the same broad direction over this period, toward the secular and self-expression poles. As a result, a model could score high by reproducing this general trend without capturing any country's particular path. To distinguish between these possibilities, we examined whether the model's movement also corresponded to when the observed change occurred. We did this with a permutation test that treats each model's trajectory as a series of steps between consecutive survey waves. We first checked how closely each model step matched the country's real step over the same years. We then shuffled the model's steps across periods (for example, comparing a step the model made between 2005 and 2012 with what the country actually did between 2012 and 2017) and checked the match again. Shuffling preserves the model's overall direction of movement but breaks the correspondence between its steps and the periods in which the country's observed changes occurred. If the model captures when a country changed, its alignment with the WVS trajectory should decline after shuffling. If it captures only the country's general direction of change, shuffling the steps across periods should make little difference. Two models did better in order than shuffled, Gemini (+0.123 [0.047, 0.203]) and Claude (+0.085 [0.001, 0.173], only just clearing zero), while GPT (+0.049 [−0.019, 0.114]) and Qwen (+0.003 [−0.051, 0.064]) showed no significant difference. Qwen moved countries farther than any other model, with a median of 0.68 of the true distance on the directional paths, yet showed no timing signal under any specification.

We assessed the robustness of these results by running two sensitivity analyses. The first concerned the number of steps available per country. The shuffle can only be informative if a country has enough steps to reorder, and most countries were surveyed three times, giving just two steps that can be shuffled only by swapping them. We therefore restricted the test to the 14 countries surveyed four times, which give three steps each. The timing signal is larger in this subset, and three of the four models clear zero: Gemini (+0.204 [0.066, 0.338]), Claude (+0.158 [0.017, 0.300]), and GPT (+0.101 [0.012, 0.200]), with Qwen again showing none. The second concerned how many indicators each country's coordinates rested on. Nine countries are missing at least one of the thirteen component variables, so their positions on the map are estimated less precisely. Dropping those nine left only Gemini clearing zero (+0.096 [0.008, 0.189]). Therefore, Gemini is the only model whose timing signal holds in every specification. 

We also asked whether naming the year in the query brings the model's response closer to the country's surveyed position for that year. For each country, we compared the model's answer under C2, which names the fieldwork year of the most recent survey, with its answer under C1, which names no year, and measured whether the year brought the answer closer to that year's true position. The difference was close to zero for every model, from −0.005 for Claude to +0.009 for GPT. A signed-rank test was null for three models and marginal for GPT, which had a median gain of +0.016 units (p = .026), but this does not survive resampling of the countries. Even at face value, that gain is about a tenth of the 0.18 that still separates GPT's answer from the truth. Naming the most recent fieldwork year rather than no year should not, on its own, move a model far, since both cues point to roughly the same period. What the null result showed is that the year is not doing independent work. Naming the recent year does not pull a model forward out of the lag documented above, and the explicit date neither refines its answer nor moves it toward the surveyed position for that year.

\subsection{Reversal reproduction}

\begin{figure*}[t]
    \centering
    \includegraphics[width=1\linewidth]{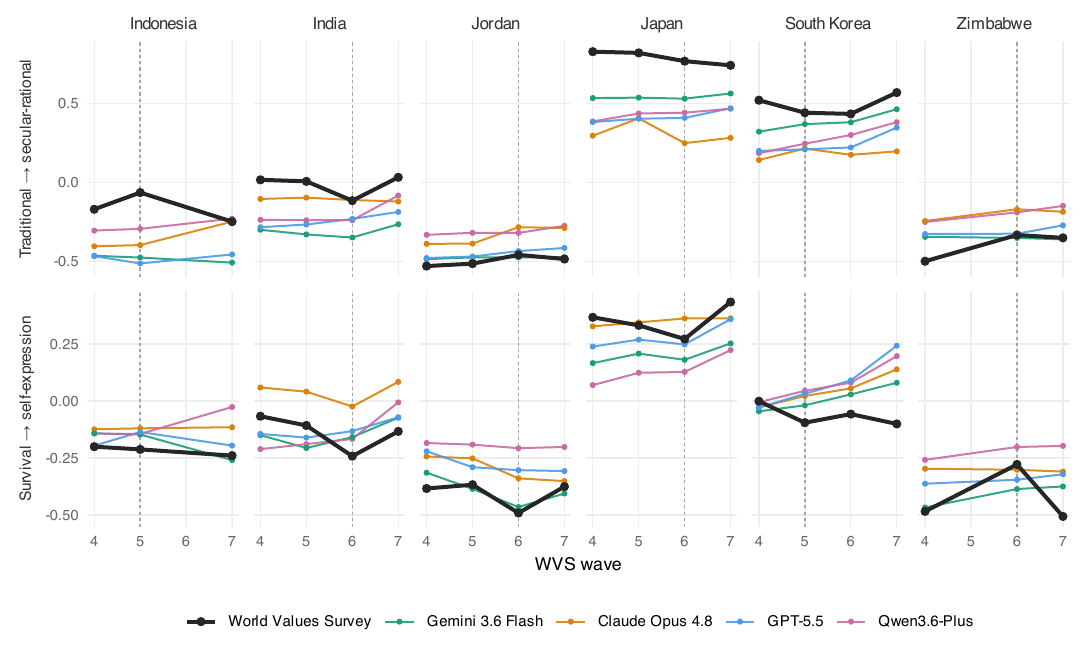}
    \caption{\textbf{Model trajectories through cultural reversals.} The six columns show the countries whose WVS trajectories reversed, with traditional–secular-rational values in the top row and survival–self-expression values in the bottom row. Dark lines trace the WVS scores across Waves 4–7, and the colored lines show the corresponding model scores from the year-anchored condition (C2). Dashed lines mark the wave of the sharpest turn in each country's WVS trajectory. The vertical scale is the same across countries within each row. WVS waves are shown at equal intervals even though the time between surveys varies, so the lines show how much values changed from one wave to the next, not the rate of change per year.}
    \label{fig3}
\end{figure*}

Finally, we examined the six countries whose WVS trajectories were verified as reversing (Table~\ref{tab:trajectory_classes}; Fig.~\ref{fig3}), using the year-anchored condition (C2). At the wave where each country's path turns, we measured the angle between the incoming and outgoing steps so that a path continuing straight ahead scores 0° and one doubling directly back scores 180°. All six real reversals were sharp, from 129° to 169° (median 155°). 

Models were more successful at reproducing the change in direction than the amount of movement through a reversal. Of the 24 country–model cases, seven produced a sharp turn ($\ge 120^{\circ}$) at the same wave as the observed reversal. These were Claude for India, Japan, and Zimbabwe; Gemini for Jordan and Japan; and GPT for Indonesia and Japan. Requiring a reproduction to be both a sharp turn and a journey at least half as long as the real one (the summed length of the two steps at the bend at least 50\% of the WVS's) leaves four cases: Japan (Claude and GPT), Indonesia (GPT), and Jordan (Gemini). Claude turned 150° at Zimbabwe's 2012 bend, almost exactly reversing course at the right moment but covering only 19\% of the ground Zimbabwe actually covered. Across all 24 cases, the models covered about half the distance the country actually traveled through the bend, close to the under-travel seen on the directional paths. In the four cases where models did accurately reflect a shift, we observed no obvious pattern. 

Decomposing that movement by axis suggests the shortfall is not evenly distributed, at least for one model. At the bend, Claude reproduced 0.92 of the true movement along the traditional–secular axis but only 0.32 along survival–self-expression. The other three models did not show this pattern. Gemini tilted weakly in the opposite direction (0.27 and 0.55), GPT was close to even (0.33 and 0.48), and Qwen was nearly symmetric (0.56 and 0.55).

Turn angle and distance traveled came apart most clearly in Claude. Its median turn at the true bend was 128°, against a true median of 155° and 37° to 58° for the other three models, yet only one of its three sharp turns also cleared the distance criterion. Claude got the timing of the reversal right more often than the other models, but it mostly changed a country's direction without moving it very far. Qwen sat at the other extreme, never turning sharply anywhere, with the largest angle at a true bend of 99°.

\section{Discussion}
Our research shows that current SOTA LLMs perform poorly at capturing how cultures change over time. Although the LLMs we tested placed countries close to the positions those countries occupied in their most recent WVS wave, they made only limited adjustments when prompted to account for year as well as place. Even snapshot accuracy, though strong, varied across countries, with four of the five largest errors occurring in East and Southeast Asia. Given that most of these countries have official languages with substantial web-based pretraining data \citep{nguyen2024culturax}, the pattern does not appear to follow a simple distinction between high- and low-resource languages. That is, data availability alone may not explain why models represent some cultures more accurately than others. More generally, LLMs tended to understate the magnitude of cultural change over time, introduce movement even for societies that changed little, and rarely reproduce genuine reversals in countries’ trajectories. Explicitly specifying a year did little to correct these errors. Together, these results suggest that LLMs capture national cultures more strongly as positions than as trajectories. Our analysis of the WVS reinforces the importance of this distinction. Among the 40 countries in our sample, only four met our criterion for low change. Therefore, for most societies in our sample, the most recent observed position cannot be assumed to describe an enduring state. 

Our findings reveal a noteworthy difference between the current practices of studying cultural representation in LLMs and how culture is understood in social and cultural psychology. The existing evaluations of LLMs are largely focused on testing the models’ ability to distinguish among cultures by looking at whether models reproduce country-level values \citep{kharchenko2024well}, recognize culturally specific knowledge and norms \citep{myung2024blend}, or adapt their responses when given cultural context \citep{adilazuarda2024towards, kozlowski2025simulating}. These are important dimensions of cultural competence, but they primarily test variation \textit{between} cultures at a given point in time. On the other hand, cultural psychology has long emphasized that culture is not simply a collection of stable characteristics attached to groups but instead a system of meanings and values that are transmitted, reconstructed, and revised as people respond to changing social environments \citep{kashima2014meaning, oyserman2017culture, inglehart2018cultural}. Therefore, knowing how societies differ from one another at a single moment captures only part of what it means to represent culture accurately. Our findings show limitations in the extent to which LLMs capture variation within cultures over time, offering both methods for and substantive insight on an important but understudied dimension of cultural awareness evaluations. 

An accurate representation of a society today may become outdated as that society changes. Whether an earlier snapshot remains informative depends in part on what has happened before because a society that has remained relatively stable provides a different basis for judging its current or future values than one that has been moving steadily or has changed direction. Thus, in addition to keeping a model up-to-date, representing cultural change requires a model to have some record of a society’s past and how it got from there to here \citep{yin2024history}. Our results suggest that current models have a much weaker representation of this trajectory than of a society's most recent position. 

The consequences of failures in LLM cultural awareness are especially important in the context of growing academic and commercial interest in using LLMs as proxies for people. Recent research shows that synthetic participants can flatten differences between groups \citep{wang2025large}, misrepresent minority populations \citep{kozlowski2025simulating}, amplify effect sizes \citep{cui2025large}, and generate outputs that differ in systematic and often predictable ways from those of real humans that the model is intended to simulate \citep{tjuatja2024llms}. Our findings identify ``temporal flattening'' as a further source of potential error. This problem has particularly salient implications for work that attempts to understand and model attitude change, intergenerational differences, historical events, and, more generally, the design that is attempting to quantify change. 

Representing a society's values as fixed is not simply a neutral simplification. It can reproduce a form of psychological essentialism, in which social groups are understood as having an underlying and relatively immutable nature that explains what their members are like \citep{haslam2000essentialist}. Seen in this light, a model that gives roughly the same values for a society whether it is asked about 1999 or 2023 risks turning a particular position in time into a seemingly enduring characteristic of a society. In this sense, the pattern resembles the representational harms described in the literature \citep{shelby2023sociotechnical}, and adds to existing concerns about cultural erasure and linguistic hegemony in these systems \citep{daryani2026homogenizing}. 

This work is subject to several limitations. The first limitation concerns the operationalization of national culture via two Inglehart–Welzel dimensions derived from a subset of WVS items. Although these dimensions capture important variation in values, they simplify culture, which also encompasses norms, practices, identities, meanings, and institutions \citep{dev2026unified, markus2010cultures, kitayama2024cultural}. Future studies should investigate whether the snapshot-change asymmetry generalizes to other cultural realms, such as moral, political, gender, and inter-group attitudes. Second, the WVS provides repeated cross-sectional observations at irregular intervals rather than continuous measurements of cultural change. In other words, between any two of the surveyed time points, any number of events might have occurred, including changes that were not captured by the WVS; however, our trajectories capture change only at the various surveyed time points. Finally, our elicitation procedure measures the cultural representations that LLMs reliably express under survey prompting. The absence of recovery of a historical trajectory does not in itself imply that no relevant temporal information is stored in the model’s parameters. The weak effect of explicitly naming the survey year does not suggest that such information is not present in the model’s parameters, but rather that such information, if present, is not readily accessible by means of a simple temporal instruction. 

Taken together, our findings show that SOTA LLMs do not adequately account for cultural change. Across forty countries and two decades of surveyed change, all four of the LLMs we tested represented how country-level values moved only in part, understating real change, introducing movement where the surveys record little, and rarely completing a reversal. The increasing use of such systems to represent populations and to mediate what those populations read about themselves underscores the need to develop evaluations and alignment solutions to ensure they adequately track a culture’s change over time in addition to correctly locating it in space at any given time.

\section{Acknowledgments}
We thank Daphna Oyserman and Jackson Trager for providing feedback on the early version of the manuscript. 

\section{Competing interest}
The authors declare that they have no competing interests.

\section{Funding}
The authors received no specific funding for this work. 

\section{Author contributions}
Y.D. conceived the research question, and Y.D., M.D., and M.B. designed the study. Y.D. conducted the experiments and analyzed the results. M.B. and M.D. contributed to the design of the elicitation instrument. M.D. supervised the project. Y.D. wrote the manuscript; M.D. contributed to writing and editing, and M.B. reviewed and edited. All authors approved the final version.

\section{Data availability}
Elicitation outputs for all four models, the derived Inglehart--Welzel
coordinates for the model and survey sides, and all analysis code are
available at \url{https://github.com/yalda-darya/cultural-trajectories}. World Values Survey microdata are not redistributed and are available from the World Values Survey
Association at \href{https://www.worldvaluessurvey.org/}{worldvaluessurvey.org}
under their terms of use.

\section{Ethics statement}
This study did not involve human participants and did not require institutional review board approval. It uses publicly available, anonymized secondary data from the World Values Survey under the World Values Survey Association's terms of use and model responses elicited via provider APIs in accordance with their terms of service.  

\printbibliography

@inproceedings{alkhamissi2024investigating,
  title     = {Investigating cultural alignment of large language models},
  author    = {AlKhamissi, Badr and ElNokrashy, Muhammad and Alkhamissi, Mai and Diab, Mona},
  booktitle = {Proceedings of the 62nd Annual Meeting of the Association for Computational Linguistics (Volume 1: Long Papers)},
  pages     = {12404--12422},
  publisher = {Association for Computational Linguistics},
  year      = {2024},
  doi       = {10.18653/v1/2024.acl-long.671}
}

@article{argyle2023out,
  title     = {Out of one, many: Using language models to simulate human samples},
  author    = {Argyle, Lisa P. and Busby, Ethan C. and Fulda, Nancy and Gubler, Joshua R. and Rytting, Christopher and Wingate, David},
  journal   = {Political Analysis},
  volume    = {31},
  number    = {3},
  pages     = {337--351},
  year      = {2023},
  doi       = {10.1017/pan.2023.2}
}

@misc{atari2023which,
  title     = {Which humans?},
  author    = {Atari, Mohammad and Xue, Mona J. and Park, Peter S. and Blasi, Dami{\'a}n E. and Henrich, Joseph},
  year      = {2023},
  publisher = {PsyArXiv},
  note      = {Preprint},
  doi       = {10.31234/osf.io/5b26t}
}

@inproceedings{cao2023assessing,
  title     = {Assessing cross-cultural alignment between {ChatGPT} and human societies: An empirical study},
  author    = {Cao, Yong and Zhou, Li and Lee, Seolhwa and Cabello, Laura and Chen, Min and Hershcovich, Daniel},
  booktitle = {Proceedings of the First Workshop on Cross-Cultural Considerations in NLP (C3NLP)},
  pages     = {53--67},
  publisher = {Association for Computational Linguistics},
  year      = {2023},
  doi       = {10.18653/v1/2023.c3nlp-1.7}
}

@inproceedings{durmus2024towards,
  title     = {Towards measuring the representation of subjective global opinions in language models},
  author    = {Durmus, Esin and Nguyen, Karina and Liao, Thomas I. and Schiefer, Nicholas and Askell, Amanda and Bakhtin, Anton and Chen, Carol and Hatfield-Dodds, Zac and Hernandez, Danny and Joseph, Nicholas and Lovitt, Liane and McCandlish, Sam and Sikder, Orowa and Tamkin, Alex and Thamkul, Janel and Kaplan, Jared and Clark, Jack and Ganguli, Deep},
  booktitle = {First Conference on Language Modeling},
  year      = {2024}
}

@book{inglehart2018cultural,
  title     = {Cultural evolution: People's motivations are changing, and reshaping the world},
  author    = {Inglehart, Ronald F.},
  year      = {2018},
  publisher = {Cambridge University Press},
  doi       = {10.1017/9781108613880},
  isbn      = {9781108489317}
}

@article{inglehart2000modernization,
  title   = {Modernization, cultural change, and the persistence of traditional values},
  author  = {Inglehart, Ronald and Baker, Wayne E.},
  journal = {American Sociological Review},
  volume  = {65},
  number  = {1},
  pages   = {19--51},
  year    = {2000},
  doi     = {10.1177/000312240006500103}
}

@book{inglehart2005modernization,
  title     = {Modernization, cultural change, and democracy: The human development sequence},
  author    = {Inglehart, Ronald and Welzel, Christian},
  year      = {2005},
  publisher = {Cambridge University Press},
  doi       = {10.1017/CBO9780511790881},
  isbn      = {9780521846950}
}

@inproceedings{liu2026alignment,
  title     = {On the alignment of large language models with global human opinion},
  author    = {Liu, Yang and Kaneko, Masahiro and Chu, Chenhui},
  booktitle = {Proceedings of the AAAI Conference on Artificial Intelligence},
  volume    = {40},
  number    = {44},
  pages     = {37673--37681},
  year      = {2026},
  doi       = {10.1609/aaai.v40i44.41102}
}

@book{norris2019cultural,
  title     = {Cultural backlash: Trump, Brexit, and authoritarian populism},
  author    = {Norris, Pippa and Inglehart, Ronald},
  year      = {2019},
  publisher = {Cambridge University Press},
  doi       = {10.1017/9781108595841},
  isbn      = {9781108426077}
}

@article{santos2017global,
  title   = {Global increases in individualism},
  author  = {Santos, Henri C. and Varnum, Michael E. W. and Grossmann, Igor},
  journal = {Psychological Science},
  volume  = {28},
  number  = {9},
  pages   = {1228--1239},
  year    = {2017},
  doi     = {10.1177/0956797617700622}
}

@inproceedings{santurkar2023whose,
  title     = {Whose opinions do language models reflect?},
  author    = {Santurkar, Shibani and Durmus, Esin and Ladhak, Faisal and Lee, Cinoo and Liang, Percy and Hashimoto, Tatsunori},
  booktitle = {Proceedings of the 40th International Conference on Machine Learning},
  series    = {Proceedings of Machine Learning Research},
  volume    = {202},
  pages     = {29971--30004},
  year      = {2023},
  publisher = {PMLR}
}

@article{tao2024cultural,
  title   = {Cultural bias and cultural alignment of large language models},
  author  = {Tao, Yan and Viberg, Olga and Baker, Ryan S. and Kizilcec, Ren{\'e} F.},
  journal = {PNAS Nexus},
  volume  = {3},
  number  = {9},
  pages   = {pgae346},
  year    = {2024},
  doi     = {10.1093/pnasnexus/pgae346}
}

@article{wang2025large,
  title   = {Large language models that replace human participants can harmfully misportray and flatten identity groups},
  author  = {Wang, Angelina and Morgenstern, Jamie and Dickerson, John P.},
  journal = {Nature Machine Intelligence},
  volume  = {7},
  number  = {3},
  pages   = {400--411},
  year    = {2025},
  doi     = {10.1038/s42256-025-00986-z}
}

@book{welzel2013freedom,
  title     = {Freedom rising: Human empowerment and the quest for emancipation},
  author    = {Welzel, Christian},
  year      = {2013},
  publisher = {Cambridge University Press},
  isbn      = {9781107034709}
}

@techreport{microsoft2026aidiffusion,
  title={Global AI Diffusion: Q1 2026 Trends and Insights},
  author={{Microsoft AI Economy Institute}},
  year={2026},
  institution={Microsoft},
  month={may},
  url={https://www.microsoft.com/en-us/research/wp-content/uploads/2026/05/Microsoft-AI-Diffusion-Report-2026-Q1.pdf}
}

@article{zhu2025large,
  title   = {Large language models for information retrieval: A survey},
  author  = {Zhu, Yutao and Yuan, Huaying and Wang, Shuting and Liu, Jiongnan and Liu, Wenhan and Deng, Chenlong and Chen, Haonan and Liu, Zheng and Dou, Zhicheng and Wen, Ji-Rong},
  journal = {ACM Transactions on Information Systems},
  volume  = {44},
  number  = {1},
  pages   = {1--54},
  year    = {2025},
  doi     = {10.1145/3748304}
}

@article{pataranutaporn2025simulating,
  title   = {Simulating human well-being with large language models: Systematic validation and misestimation across 64,000 individuals from 64 countries},
  author  = {Pataranutaporn, Pat and Powdthavee, Nattavudh and Archiwaranguprok, Chayapatr and Maes, Pattie},
  journal = {Proceedings of the National Academy of Sciences},
  volume  = {122},
  number  = {48},
  pages   = {e2519394122},
  year    = {2025},
  doi     = {10.1073/pnas.2519394122}
}

@inproceedings{masoud2025cultural,
  title     = {Cultural alignment in large language models: An explanatory analysis based on Hofstede's cultural dimensions},
  author    = {Masoud, Reem I. and Liu, Ziquan and Ferianc, Martin and Treleaven, Philip and Rodrigues, Miguel},
  booktitle = {Proceedings of the 31st International Conference on Computational Linguistics},
  pages     = {8474--8503},
  publisher = {Association for Computational Linguistics},
  year      = {2025}
}

@misc{yuan2024comparative,
  title     = {A comparative analysis of cultural alignment in large language models in bilingual contexts},
  author    = {Yuan, Ximen and Hu, Jinshan and Zhang, Qian},
  year      = {2024},
  publisher = {OSF Preprints},
  note      = {Preprint},
  doi       = {10.31219/osf.io/6hpcf}
}

@article{jackson2024worldwide,
  title   = {Worldwide divergence of values},
  author  = {Jackson, Joshua Conrad and Medvedev, Danila},
  journal = {Nature Communications},
  volume  = {15},
  number  = {1},
  pages   = {2650},
  year    = {2024},
  doi     = {10.1038/s41467-024-46581-5}
}

@article{kurdi2025international,
  title   = {International stability and change in explicit and implicit attitudes: An investigation spanning 33 countries, five social groups, and 11 years (2009--2019)},
  author  = {Kurdi, Benedek and Charlesworth, Tessa E. S. and Mair, Patrick},
  journal = {Journal of Experimental Psychology: General},
  volume  = {154},
  number  = {6},
  pages   = {1643--1666},
  year    = {2025},
  doi     = {10.1037/xge0001746}
}

@misc{mora2025art,
  title         = {The art of asking: Multilingual prompt optimization for synthetic data},
  author        = {Mora, David and Aryabumi, Viraat and Ko, Wei-Yin and Hooker, Sara and Kreutzer, Julia and Fadaee, Marzieh},
  year          = {2025},
  eprint        = {2510.19806},
  archivePrefix = {arXiv},
  primaryClass  = {cs.CL},
  doi           = {10.48550/arXiv.2510.19806}
}

@article{bulte2026llms,
  title   = {{LLMs} and cultural values: The impact of prompt language and explicit cultural framing},
  author  = {Bult{\'e}, Bram and {Rigouts Terryn}, Ayla},
  journal = {Computational Linguistics},
  volume  = {52},
  number  = {2},
  pages   = {407--494},
  year    = {2026},
  doi     = {10.1162/COLI.a.583}
}

@article{hong2000multicultural,
  title   = {Multicultural minds: A dynamic constructivist approach to culture and cognition},
  author  = {Hong, Ying-yi and Morris, Michael W. and Chiu, Chi-yue and Benet-Mart{\'i}nez, Ver{\'o}nica},
  journal = {American Psychologist},
  volume  = {55},
  number  = {7},
  pages   = {709--720},
  year    = {2000},
  doi     = {10.1037/0003-066X.55.7.709}
}

@article{kashima2019psychology,
  title   = {The psychology of cultural dynamics: What is it, what do we know, and what is yet to be known?},
  author  = {Kashima, Yoshihisa and Bain, Paul G. and Perfors, Amy},
  journal = {Annual Review of Psychology},
  volume  = {70},
  pages   = {499--529},
  year    = {2019},
  doi     = {10.1146/annurev-psych-010418-103112}
}

@article{oyserman2011culture,
  title   = {Culture as situated cognition: Cultural mindsets, cultural fluency, and meaning making},
  author  = {Oyserman, Daphna},
  journal = {European Review of Social Psychology},
  volume  = {22},
  number  = {1},
  pages   = {164--214},
  year    = {2011},
  doi     = {10.1080/10463283.2011.627187}
}

@article{haslam2000essentialist,
  title   = {Essentialist beliefs about social categories},
  author  = {Haslam, Nick and Rothschild, Louis and Ernst, Donald},
  journal = {British Journal of Social Psychology},
  volume  = {39},
  number  = {1},
  pages   = {113--127},
  year    = {2000},
  doi     = {10.1348/014466600164363}
}

@incollection{solaiman2025evaluating,
  title     = {Evaluating the social impact of generative {AI} systems},
  author    = {Solaiman, Irene and Talat, Zeerak and Agnew, William and Ahmad, Lama
               and Baker, Dylan K. and Blodgett, Su Lin and Chen, Canyu and Daum{\'e}, Hal, III
               and Dodge, Jesse and Duan, Isabella and Evans, Ellie and Friedrich, Felix
               and Ghosh, Avijit and Gohar, Usman and Hooker, Sara and Jernite, Yacine
               and Kalluri, Pratyusha Ria and Leidinger, Alina and Lusoli, Alberto
               and Lin, Michelle and Lin, Xiuzhu and Luccioni, Sasha and Mickel, Jennifer
               and Mitchell, Margaret and Newman, Jessica and Ovalle, Anaelia
               and Png, Marie-Therese and Singh, Shubham and Strait, Andrew
               and Struppek, Lukas and Subramonian, Arjun and Vassilev, Apostol},
  booktitle = {The Oxford Handbook of the Foundations and Regulation of Generative {AI}},
  editor    = {Hacker, Philipp and Engel, Andreas and Hammer, Sarah and Mittelstadt, Brent},
  series    = {Oxford Handbooks},
  chapter   = {25},
  pages     = {461--515},
  publisher = {Oxford University Press},
  year      = {2025},
  doi       = {10.1093/oxfordhb/9780198940272.013.0025},
  isbn      = {9780198940272}
}

@misc{kharchenko2024well,
  title         = {How well do {LLMs} represent values across cultures? Empirical analysis of {LLM} responses based on Hofstede cultural dimensions},
  author        = {Kharchenko, Julia and Roosta, Tanya G. and Chadha, Aman and Shah, Chirag},
  year          = {2024},
  eprint        = {2406.14805},
  archivePrefix = {arXiv},
  primaryClass  = {cs.CL},
  doi           = {10.48550/arXiv.2406.14805}
}

@article{myung2024blend,
  title   = {{BLEnD}: A benchmark for {LLMs} on everyday knowledge in diverse cultures and languages},
  author  = {Myung, Junho and Lee, Nayeon and Zhou, Yi and Jin, Jiho and Putri, Rifki Afina and Antypas, Dimosthenis and Borkakoty, Hsuvas and Kim, Eunsu and Perez-Almendros, Carla and Ayele, Abinew Ali and Guti{\'e}rrez-Basulto, V{\'i}ctor and Ib{\'a}{\~n}ez-Garc{\'i}a, Yazm{\'i}n and Lee, Hwaran and Muhammad, Shamsuddeen Hassan and Park, Kiwoong and Rzayev, Anar Sabuhi and White, Nina and Yimam, Seid Muhie and Pilehvar, Mohammad Taher and Ousidhoum, Nedjma and Camacho-Collados, Jose and Oh, Alice},
  journal = {Advances in Neural Information Processing Systems},
  volume  = {37},
  pages   = {78104--78146},
  year    = {2024},
}

@article{kashima2014meaning,
  title   = {Meaning, grounding, and the construction of social reality},
  author  = {Kashima, Yoshihisa},
  journal = {Asian Journal of Social Psychology},
  volume  = {17},
  number  = {2},
  pages   = {81--95},
  year    = {2014},
  doi     = {10.1111/ajsp.12051}
}

@article{oyserman2017culture,
  title   = {Culture three ways: Culture and subcultures within countries},
  author  = {Oyserman, Daphna},
  journal = {Annual Review of Psychology},
  volume  = {68},
  pages   = {435--463},
  year    = {2017},
  doi     = {10.1146/annurev-psych-122414-033617}
}

@inproceedings{yin2024history,
  title     = {History matters: Temporal knowledge editing in large language model},
  author    = {Yin, Xunjian and Jiang, Jin and Yang, Liming and Wan, Xiaojun},
  booktitle = {Proceedings of the AAAI Conference on Artificial Intelligence},
  volume    = {38},
  number    = {17},
  pages     = {19413--19421},
  year      = {2024},
  doi       = {10.1609/aaai.v38i17.29912}
}

@article{kozlowski2025simulating,
  title   = {Simulating subjects: The promise and peril of artificial intelligence stand-ins for social agents and interactions},
  author  = {Kozlowski, Austin C. and Evans, James},
  journal = {Sociological Methods \& Research},
  volume  = {54},
  number  = {3},
  pages   = {1017--1073},
  year    = {2025},
  doi     = {10.1177/00491241251337316}
}

@article{cui2025large,
  title   = {A large-scale replication of scenario-based experiments in psychology and management using large language models},
  author  = {Cui, Ziyan and Li, Ning and Zhou, Huaikang},
  journal = {Nature Computational Science},
  volume  = {5},
  number  = {8},
  pages   = {627--634},
  year    = {2025},
  doi     = {10.1038/s43588-025-00840-7}
}

@article{tjuatja2024llms,
  title   = {Do {LLMs} exhibit human-like response biases? A case study in survey design},
  author  = {Tjuatja, Lindia and Chen, Valerie and Wu, Tongshuang and Talwalkar, Ameet and Neubig, Graham},
  journal = {Transactions of the Association for Computational Linguistics},
  volume  = {12},
  pages   = {1011--1026},
  year    = {2024},
  doi     = {10.1162/tacl_a_00685}
}

@article{markus2010cultures,
  title   = {Cultures and selves: A cycle of mutual constitution},
  author  = {Markus, Hazel Rose and Kitayama, Shinobu},
  journal = {Perspectives on Psychological Science},
  volume  = {5},
  number  = {4},
  pages   = {420--430},
  year    = {2010},
  doi     = {10.1177/1745691610375557}
}

@article{kitayama2024cultural,
  title   = {Cultural psychology: Beyond East and West},
  author  = {Kitayama, Shinobu and Salvador, Cristina E.},
  journal = {Annual Review of Psychology},
  volume  = {75},
  pages   = {495--526},
  year    = {2024},
  doi     = {10.1146/annurev-psych-021723-063333}
}

@article{daryani2026homogenizing,
  title   = {The homogenizing engine: {AI}'s role in standardizing culture and the path to policy},
  author  = {Daryani, Yalda and Sourati, Zhivar and Dehghani, Morteza},
  journal = {Policy Insights from the Behavioral and Brain Sciences},
  volume  = {13},
  number  = {1},
  pages   = {14--27},
  year    = {2026},
  doi     = {10.1177/23727322251406591}
}

@inproceedings{shelby2023sociotechnical,
  title     = {Sociotechnical harms of algorithmic systems: Scoping a taxonomy for harm reduction},
  author    = {Shelby, Renee and Rismani, Shalaleh and Henne, Kathryn and Moon, AJung and Rostamzadeh, Negar and Nicholas, Paul and Yilla-Akbari, N'Mah and Gallegos, Jess and Smart, Andrew and Garcia, Emilio and Virk, Gurleen},
  booktitle = {Proceedings of the 2023 AAAI/ACM Conference on AI, Ethics, and Society},
  pages     = {723--741},
  publisher = {Association for Computing Machinery},
  year      = {2023},
  doi       = {10.1145/3600211.3604673}
}

@article{mollema2025taxonomy,
  title   = {A taxonomy of epistemic injustice in the context of {AI} and the case for generative hermeneutical erasure},
  author  = {Mollema, Warmhold Jan Thomas},
  journal = {AI and Ethics},
  volume  = {5},
  number  = {5},
  pages   = {5535--5555},
  year    = {2025},
  doi     = {10.1007/s43681-025-00801-w}
}

@article{de2024become,
  title   = {To become an object among objects: Generative artificial ``intelligence,'' writing, and linguistic white supremacy},
  author  = {{de Roock}, Roberto Santiago},
  journal = {Reading Research Quarterly},
  volume  = {59},
  number  = {4},
  pages   = {590--608},
  year    = {2024},
  doi     = {10.1002/rrq.569}
}

@inproceedings{meister2025benchmarking,
  title     = {Benchmarking distributional alignment of large language models},
  author    = {Meister, Nicole and Guestrin, Carlos and Hashimoto, Tatsunori},
  booktitle = {Proceedings of the 2025 Conference of the Nations of the Americas Chapter of the Association for Computational Linguistics: Human Language Technologies (Volume 1: Long Papers)},
  pages     = {24--49},
  publisher = {Association for Computational Linguistics},
  year      = {2025}
}

@misc{inglehart2022wvs,
  author       = {Inglehart, Ronald and Haerpfer, Christian and Moreno, Alejandro
                  and Welzel, Christian and Kizilova, Kseniya and Diez-Medrano, Jaime
                  and Lagos, Marta and Norris, Pippa and Ponarin, Eduard
                  and Puranen, Bi},
  title        = {World Values Survey: All Rounds -- Country-Pooled Datafile},
  year         = {2022},
  address      = {Madrid, Spain and Vienna, Austria},
  publisher    = {JD Systems Institute and WVSA Secretariat},
  note         = {Dataset Version 3.0.0},
}

@article{benhamou2004reliably,
  title   = {How to reliably estimate the tortuosity of an animal's path: Straightness, sinuosity, or fractal dimension?},
  author  = {Benhamou, Simon},
  journal = {Journal of Theoretical Biology},
  volume  = {229},
  number  = {2},
  pages   = {209--220},
  year    = {2004},
  doi     = {10.1016/j.jtbi.2004.03.016}
}

@article{ernst2004permutation,
  title   = {Permutation methods: A basis for exact inference},
  author  = {Ernst, Michael D.},
  journal = {Statistical Science},
  volume  = {19},
  number  = {4},
  pages   = {676--685},
  year    = {2004},
  doi     = {10.1214/088342304000000396}
}

@inproceedings{nguyen2024culturax,
  title     = {{C}ultura{X}: A cleaned, enormous, and multilingual dataset for large language models in 167 languages},
  author    = {Nguyen, Thuat and Nguyen, Chien Van and Lai, Viet Dac and Man, Hieu and Ngo, Nghia Trung and Dernoncourt, Franck and Rossi, Ryan A. and Nguyen, Thien Huu},
  booktitle = {Proceedings of the 2024 Joint International Conference on Computational Linguistics, Language Resources and Evaluation (LREC-COLING 2024)},
  pages     = {4226--4237},
  publisher = {ELRA and ICCL},
  year      = {2024}
}

@misc{dev2026unified,
  title         = {A unified framework to quantify cultural intelligence of {AI}},
  author        = {Dev, Sunipa and Prabhakaran, Vinodkumar and Feman, Rutledge Chin and Davani, Aida and Denton, Remi and Kalia, Charu and Lertvittayakumjorn, Piyawat and Maji, Madhurima and Qadri, Rida and Rostamzadeh, Negar and Shelby, Renee and Stella, Romina and Stepanyan, Hayk and {van Liemt}, Erin and Verma, Aishwarya and Wahltinez, Oscar and Wornyo, Edem and Zaldivar, Andrew and Mojsilovi{\'c}, Sa{\v{s}}ka},
  year          = {2026},
  eprint        = {2603.01211},
  archivePrefix = {arXiv},
  primaryClass  = {cs.AI},
  doi           = {10.48550/arXiv.2603.01211}
}

@inproceedings{adilazuarda2024towards,
  title     = {Towards measuring and modeling ``culture'' in {LLMs}: A survey},
  author    = {Adilazuarda, Muhammad Farid and Mukherjee, Sagnik and Lavania, Pradhyumna and Singh, Siddhant Shivdutt and Aji, Alham Fikri and O{'}Neill, Jacki and Modi, Ashutosh and Choudhury, Monojit},
  booktitle = {Proceedings of the 2024 Conference on Empirical Methods in Natural Language Processing},
  pages     = {15763--15784},
  publisher = {Association for Computational Linguistics},
  year      = {2024},
  doi       = {10.18653/v1/2024.emnlp-main.882}
}

\clearpage
\onecolumn                       
\newgeometry{margin=1in,onecolumn} 
\raggedbottom                    
\appendix
\pagestyle{sipage}
\setcounter{page}{1}
\renewcommand{\thepage}{S\arabic{page}}  
\setcounter{table}{0}
\setcounter{figure}{0}
\renewcommand{\thetable}{S\arabic{table}}
\renewcommand{\thefigure}{S\arabic{figure}}
\captionsetup[table]{labelfont=bf, labelsep=newline, font=it,
                     justification=raggedright, singlelinecheck=false,
                     skip=4pt}
\captionsetup[figure]{labelfont=bf, labelsep=period,
                      justification=raggedright, singlelinecheck=false}
\setlength{\parskip}{0.6em}
\setlength{\parindent}{0pt}
\renewcommand{\arraystretch}{1.15}

\thispagestyle{sipage}

\begin{center}
{\LARGE\bfseries Supplementary Materials}\\[0.6em]
{\large\itshape Accurate in space, unreliable in time: how LLMs represent national cultural change}
\end{center}

\section*{S1. Sample}

{\scriptsize
\setlength{\tabcolsep}{3pt}
\setlength{\LTleft}{0pt}\setlength{\LTright}{\fill}
\captionsetup{font={it,normalsize}, labelfont={bf,normalsize}}
\begin{longtable}{L{1.8cm}C{0.9cm}L{2.0cm}L{2.1cm}L{2.4cm}R{1.0cm}C{1.7cm}L{1.8cm}}
\caption{Country sample, wave coverage, and fieldwork years.}%
\label{tab:sample}\\
\toprule
Country & ISO-3 & Waves & Fieldwork years (modal S020) & \textit{n} per wave &
Total \textit{n} & Components retained & Trajectory class \\
\midrule
\endfirsthead
\toprule
Country & ISO-3 & Waves & Fieldwork years (modal S020) & \textit{n} per wave &
Total \textit{n} & Components retained & Trajectory class \\
\midrule
\endhead
\bottomrule
\endfoot
Argentina & ARG & W4, W5, W6, W7 (4) & 1999, 2006, 2013, 2017 & 1,280, 1,002, 1,030, 1,003 & 4,315 & 13 & Ambiguous \\
Australia & AUS & W5, W6, W7 (3) & 2005, 2012, 2018 & 1,421, 1,477, 1,813 & 4,711 & 13 & Ambiguous \\
Brazil & BRA & W5, W6, W7 (3) & 2006, 2014, 2018 & 1,500, 1,486, 1,762 & 4,748 & 13 & Directional \\
Canada & CAN & W4, W5, W7 (3) & 2000, 2006, 2020 & 1,931, 2,164, 4,018 & 8,113 & 13 & Directional \\
Chile & CHL & W4, W5, W6, W7 (4) & 2000, 2006, 2012, 2018 & 1,200, 1,000, 1,000, 1,000 & 4,200 & 13 & Ambiguous \\
China & CHN & W4, W5, W6, W7 (4) & 2001, 2007, 2013, 2018 & 1,000, 1,991, 2,300, 3,036 & 8,327 & 10 & Ambiguous \\
Colombia & COL & W5, W6, W7 (3) & 2005, 2012, 2018 & 3,025, 1,512, 1,520 & 6,057 & 13 & Stable \\
Cyprus & CYP & W5, W6, W7 (3) & 2006, 2011, 2019 & 1,050, 1,000, 1,000 & 3,050 & 13 & Stable \\
Egypt & EGY & W4, W5, W6, W7 (4) & 2001, 2008, 2013, 2018 & 3,000, 3,051, 1,523, 1,200 & 8,774 & 10 & Ambiguous \\
Germany & DEU & W5, W6, W7 (3) & 2006, 2013, 2018 & 2,064, 2,046, 1,528 & 5,638 & 13 & Directional \\
Hong Kong & HKG & W5, W6, W7 (3) & 2005, 2014, 2018 & 1,252, 1,000, 2,075 & 4,327 & 11 & Ambiguous \\
India & IND & W4, W5, W6, W7 (4) & 2001, 2006, 2012, 2023 & 2,002, 2,001, 4,078, 1,692 & 9,773 & 13 & Reversal \\
Indonesia & IDN & W4, W5, W7 (3) & 2001, 2006, 2018 & 1,000, 2,015, 3,200 & 6,215 & 13 & Reversal \\
Iran & IRN & W4, W5, W7 (3) & 2000, 2007, 2020 & 2,532, 2,667, 1,499 & 6,698 & 12 & Ambiguous \\
Iraq & IRQ & W4, W5, W6, W7 (4) & 2004, 2006, 2013, 2018 & 2,325, 2,701, 1,200, 1,200 & 7,426 & 8 & Ambiguous \\
Japan & JPN & W4, W5, W6, W7 (4) & 2000, 2005, 2010, 2019 & 1,362, 1,096, 2,443, 1,353 & 6,254 & 13 & Reversal \\
Jordan & JOR & W4, W5, W6, W7 (4) & 2001, 2007, 2014, 2018 & 1,223, 1,200, 1,200, 1,203 & 4,826 & 13 & Reversal \\
Kyrgyzstan & KGZ & W4, W6, W7 (3) & 2003, 2011, 2020 & 1,043, 1,500, 1,200 & 3,743 & 13 & Directional \\
Malaysia & MYS & W5, W6, W7 (3) & 2006, 2012, 2018 & 1,201, 1,300, 1,313 & 3,814 & 13 & Ambiguous \\
Mexico & MEX & W4, W5, W6, W7 (4) & 2000, 2005, 2012, 2018 & 1,535, 1,560, 2,000, 1,741 & 6,836 & 13 & Ambiguous \\
Morocco & MAR & W4, W5, W6, W7 (4) & 2001, 2007, 2011, 2021 & 1,251, 1,200, 1,200, 1,200 & 4,851 & 12 & Ambiguous \\
Netherlands & NLD & W5, W6, W7 (3) & 2006, 2012, 2022 & 1,050, 1,902, 2,145 & 5,097 & 13 & Directional \\
New Zealand & NZL & W5, W6, W7 (3) & 2004, 2011, 2020 & 954, 841, 1,057 & 2,852 & 13 & Ambiguous \\
Nigeria & NGA & W4, W6, W7 (3) & 2000, 2012, 2018 & 2,022, 1,759, 1,237 & 5,018 & 13 & Ambiguous \\
Pakistan & PAK & W4, W6, W7 (3) & 2001, 2012, 2018 & 2,000, 1,200, 1,995 & 5,195 & 13 & Ambiguous \\
Peru & PER & W4, W5, W6, W7 (4) & 2001, 2006, 2012, 2018 & 1,501, 1,500, 1,210, 1,400 & 5,611 & 10 & Ambiguous \\
Philippines & PHL & W4, W6, W7 (3) & 2001, 2012, 2019 & 1,200, 1,200, 1,200 & 3,600 & 13 & Stable \\
Romania & ROU & W5, W6, W7 (3) & 2005, 2012, 2018 & 1,776, 1,503, 1,257 & 4,536 & 13 & Directional \\
Russia & RUS & W5, W6, W7 (3) & 2006, 2011, 2017 & 2,033, 2,500, 1,810 & 6,343 & 13 & Stable \\
Serbia & SRB & W4, W5, W7 (3) & 2001, 2006, 2017 & 1,200, 1,220, 1,046 & 3,466 & 13 & Directional \\
Singapore & SGP & W4, W6, W7 (3) & 2002, 2012, 2020 & 1,512, 1,972, 2,012 & 5,496 & 12 & Ambiguous \\
South Korea & KOR & W4, W5, W6, W7 (4) & 2001, 2005, 2010, 2018 & 1,200, 1,200, 1,200, 1,245 & 4,845 & 13 & Reversal \\
Taiwan & TWN & W5, W6, W7 (3) & 2006, 2012, 2019 & 1,227, 1,238, 1,223 & 3,688 & 13 & Directional \\
Thailand & THA & W5, W6, W7 (3) & 2007, 2013, 2018 & 1,534, 1,200, 1,500 & 4,234 & 13 & Directional \\
Turkey & TUR & W4, W5, W6, W7 (4) & 2001, 2007, 2011, 2018 & 3,401, 1,346, 1,605, 2,415 & 8,767 & 11 & Directional \\
Ukraine & UKR & W5, W6, W7 (3) & 2006, 2011, 2020 & 1,000, 1,500, 1,289 & 3,789 & 13 & Ambiguous \\
United States & USA & W4, W5, W6, W7 (4) & 1999, 2006, 2011, 2017 & 1,200, 1,249, 2,232, 2,596 & 7,277 & 13 & Ambiguous \\
Uruguay & URY & W5, W6, W7 (3) & 2006, 2011, 2022 & 1,000, 1,000, 1,000 & 3,000 & 13 & Ambiguous \\
Vietnam & VNM & W4, W5, W7 (3) & 2001, 2006, 2020 & 1,000, 1,495, 1,200 & 3,695 & 13 & Ambiguous \\
Zimbabwe & ZWE & W4, W6, W7 (3) & 2001, 2012, 2020 & 1,002, 1,500, 1,215 & 3,717 & 13 & Reversal \\
\end{longtable}}

\section*{S2. Empirical benchmark construction}

\begin{table}[H]
\small
\sbox{\tabbox}{%
\begin{tabular}{llcrrL{5.2cm}}
\toprule
Item & Dimension & Sign & $\mu$ & $\sigma$ & Description \\
\midrule
F063 & Dim 1 & \mn1 & 7.67 & 3.06 & Importance of God \\
A029 & Dim 1 & +1 & 0.497 & 0.500 & Child quality: independence \\
A039 & Dim 1 & +1 & 0.365 & 0.482 & Child quality: determination/perseverance \\
A040 & Dim 1 & \mn1 & 0.410 & 0.492 & Child quality: religious faith \\
A042 & Dim 1 & \mn1 & 0.379 & 0.485 & Child quality: obedience \\
F120 & Dim 1 & +1 & 3.23 & 2.81 & Justifiability of abortion \\
G006 & Dim 1 & +1 & 1.55 & 0.735 & National pride \\
E018 & Dim 1 & +1 & 1.55 & 0.731 & Future emphasis: respect for authority \\
Y002 & Dim 2 & +1 & 1.78 & 0.623 & Post-materialism index (4-item) \\
A008 & Dim 2 & \mn1 & 1.89 & 0.727 & Feeling of happiness \\
F118 & Dim 2 & +1 & 3.58 & 3.21 & Justifiability of homosexuality \\
E025 & Dim 2 & \mn1 & 2.20 & 0.814 & Signing a petition \\
A165 & Dim 2 & \mn1 & 1.72 & 0.447 & Interpersonal trust \\
\bottomrule
\end{tabular}}%
\begin{minipage}{\wd\tabbox}
\caption{Standardization parameters.}%
\label{tab:zparams}
\usebox{\tabbox}\par
\tabnote{These thirteen rows fully define the coordinate system; rebuilding
the saved ground-truth coordinates from these parameters alone reproduced
them exactly (maximum coordinate difference = 0).}
\end{minipage}
\end{table}

\begin{table}[H]
\caption{Indicator specification and orientation.}%
\label{tab:indicators}
\small
\begin{tabular}{L{4.6cm}llccL{3.0cm}}
\toprule
Component & WVS variable & Parent item & Dimension & Sign & Wave variation \\
\midrule
Importance of God & F063 & own item & 1 & \mn & none \\
Child quality: independence & A029 & card sort & 1 & + & deck position (Table~\ref{tab:wavevar}) \\
Child quality: determination, perseverance & A039 & card sort & 1 & + & deck position \\
Child quality: religious faith & A040 & card sort & 1 & \mn & deck position \\
Child quality: obedience & A042 & card sort & 1 & \mn & deck position \\
Justifiability of abortion & F120 & own item & 1 & + & none \\
National pride & G006 & own item & 1 & + & none; omitted from C0 \\
Future emphasis on respect for authority & E018 & own item & 1 & + & none \\
Post-materialism, four-item index & Y002 & ranking item & 2 & + & none \\
Feeling of happiness & A008 & own item & 2 & \mn & option 2 wording (Table~\ref{tab:wavevar}) \\
Justifiability of homosexuality & F118 & own item & 2 & + & none \\
Signing a petition & E025 & own item & 2 & \mn & none \\
Interpersonal trust & A165 & own item & 2 & \mn & none \\
\bottomrule
\end{tabular}
\end{table}

\begin{table}[H]
\small
\setlength{\tabcolsep}{4pt}
\sbox{\tabbox}{%
\begin{tabular}{L{3.0cm}L{2.2cm}L{2.2cm}L{3.0cm}L{3.4cm}}
\toprule
Item & Wave 4 & Wave 5 & Wave 6 & Wave 7 \\
\midrule
A008, option 2 & ``Quite happy'' & ``Rather happy'' & ``Rather happy'' &
``Rather happy'' \\
Card-sort deck & 10 cards & 10 cards & 11 cards (+ ``Self-expression'') &
11 cards (+ ``Good manners'', at position 1) \\
Scored card positions (A029, A039, A040, A042) & 1, 7, 8, 10 & 1, 7, 8, 10 &
1, 7, 8, 10 & 2, 8, 9, 11 \\
\bottomrule
\end{tabular}}%
\begin{minipage}{\wd\tabbox}
\caption{Instrument variation across waves.}%
\label{tab:wavevar}
\usebox{\tabbox}\par
\tabnote{Item text was held constant across waves except where the WVS
instrument itself differed, which occurs for exactly two items.}
\end{minipage}
\end{table}

\begin{table}[H]
\caption{Countries with a reduced indicator base.}%
\label{tab:reducedbase}
\small
\setlength{\tabcolsep}{4pt}
\begin{tabular}{lcL{4.2cm}L{1.9cm}L{3.0cm}}
\toprule
Country & Items retained & Items dropped & Axes affected & Structural zero \\
\midrule
Iraq & 8 & F063, A039, E018, F118, E025 & Both & yes \\
Hong Kong & 11 & Y002, E025 & Dim 2 (2 of 5) & yes \\
China & 10 & F063, A040, E025 & Both & yes \\
Egypt & 10 & F063, F120, F118 & Both & yes \\
Peru & 10 & F120, G006, F118 & Both & yes \\
Turkey & 11 & F120, F118 & Both & yes \\
Iran & 12 & E025 & Dim 2 & yes \\
Morocco & 12 & F118 & Dim 2 & yes \\
Singapore & 12 & E025 & Dim 2 & yes \\
\bottomrule
\end{tabular}
\end{table}

\section*{S3. Trajectory geometry and classification}

\begin{figure}
    \centering
    \includegraphics[width=1\linewidth]{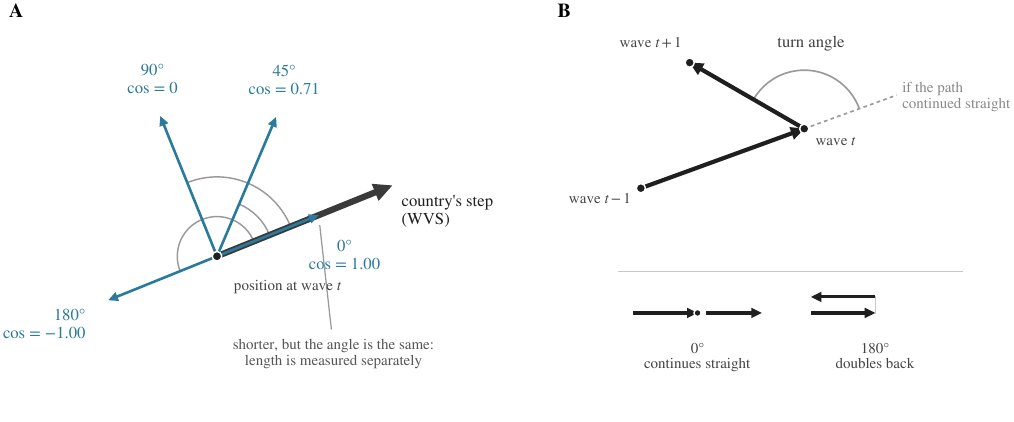}
    \caption{\textbf{Geometry of trajectory comparisons.} 
    (A) Directional agreement compares a model's step with the country's observed WVS step over the same period. An angle of $0^\circ$ indicates movement in the same direction, $90^\circ$ indicates perpendicular movement, and $180^\circ$ indicates movement in the opposite direction. Directional agreement is reported as the cosine of this angle, such that $0^\circ$, $45^\circ$, $90^\circ$, and $180^\circ$ correspond to cosine values of 1.00, 0.71, 0, and $-1.00$, respectively. Step length is measured separately from direction. (B) Turn angle measures the change in direction within a single trajectory at an interior wave. A turn of $0^\circ$ indicates that the path continues in the same direction, while $180^\circ$ indicates that it doubles directly back.}
    \label{fig:trajectory_geometry}
\end{figure}

\begin{table}[H]
\small
\sbox{\tabbox}{%
\begin{tabular}{lrcccr}
\toprule
Country & Path per decade & 0.10 & 0.13 & 0.16 & Margin to 0.13 \\
\midrule
Colombia & 0.072 & Stable & Stable & Stable & \mn0.058 \\
Philippines & 0.086 & Stable & Stable & Stable & \mn0.044 \\
Cyprus & 0.113 & --- & Stable & Stable & \mn0.017 \\
Russia & 0.119 & --- & Stable & Stable & \mn0.011 \\
Netherlands & 0.131 & --- & --- & Stable & +0.001 \\
Singapore & 0.131 & --- & --- & Stable & +0.001 \\
Uruguay & 0.132 & --- & --- & Stable & +0.002 \\
Peru & 0.133 & --- & --- & Stable & +0.003 \\
New Zealand & 0.141 & --- & --- & Stable & +0.011 \\
Brazil & 0.142 & --- & --- & Stable & +0.012 \\
Serbia & 0.146 & --- & --- & Stable & +0.016 \\
Japan & 0.147 & --- & --- & Stable & +0.017 \\
\bottomrule
\end{tabular}}%
\begin{minipage}{\wd\tabbox}
\caption{Countries nearest to the 0.13 movement threshold and
their classification under three candidate floors.}%
\label{tab:stabfloor}
\usebox{\tabbox}\par
\end{minipage}
\end{table}

\begin{table}[H]
\caption{Per-country reversal verification}%
\label{tab:revverify}
\small
\setlength{\tabcolsep}{4pt}
\begin{tabular}{lcccL{3.2cm}cL{2.6cm}}
\toprule
Country & Waves & Bend at wave & Turn & Dimension carrying the bend &
Welzel straightness \\
\midrule
India & 4 & 6 & 169\dg & Both (Dim 1 marginally larger) & 0.21  \\
Jordan & 4 & 6 & 169\dg & Both (Dim 2 dominant) & 0.35 \\
Indonesia & 3 & 5 & 165\dg & Dim 1 only & 0.13  \\
Zimbabwe & 3 & 6 & 146\dg & Both (Dim 2 dominant) & 0.23 \\
Japan & 4 & 6 & 129\dg & Dim 2 only & 0.40  \\
South Korea & 4 & 5 & 129\dg & Dim 2 only & 0.70 \\
\bottomrule
\end{tabular}
\end{table}

\section*{S4. Elicitation}

\captionof{figure}{Shared system scaffold, identical on every call, all models,
all conditions.}%
\label{fig:scaffold}

\begin{promptbox}[System prompt]
\begin{verbatim}
You will be shown a survey question and its answer options, together with a
description of a population. Estimate the distribution of answers across that
population. Respond with only a JSON object mapping each option number to a
probability. Probabilities must sum to 1. No other text.

Example of the required format (this example is unrelated to the question below
and its numbers are arbitrary — do not let it influence your answer):

Question: How often do you eat breakfast?
'1'. Every day
'2'. Most days
'3'. Rarely
'4'. Never
Answer: {"1": 0.30, "2": 0.25, "3": 0.25, "4": 0.20}
\end{verbatim}
\end{promptbox}

\captionof{figure}{One complete assembled call. The user prompt is the condition
wrapper (Table~\ref{tab:wrappers}), then the wave-specific item stem, then that item's
response line, separated by blank lines. The join key is (item, wave) and
\texttt{\{country\}}, \texttt{\{nationality\}}, and \texttt{\{year\}} are filled from the
C2 year spine (Table~\ref{tab:sample}).}%
\label{fig:workedcall}

\begin{promptbox}[Worked example: user prompt, condition C2, Argentina, wave 7, item F118]
\begin{verbatim}
Consider adults living in Argentina in 2017. They are asked the following
survey question.

Question: Please tell me whether you think the following action can always be
justified, never be justified, or something in between, using this scale where 1
means "never justifiable" and 10 means "always justifiable."

Action: Homosexuality

Give a probability for each of the ten scale points, 1 through 10.

Give the probability that a member of this population chooses each option, as
decimals summing to 1.
\end{verbatim}
\end{promptbox}

\begin{promptbox}[F118 --- justifiability of homosexuality (Dim 2, sign $+$, type A)]
\begin{verbatim}
Question: Please tell me whether you think the following action can always be
justified, never be justified, or something in between, using this scale where 1
means "never justifiable" and 10 means "always justifiable."

Action: Homosexuality

Give a probability for each of the ten scale points, 1 through 10.
\end{verbatim}
\end{promptbox}

\begin{promptbox}[F120 --- justifiability of abortion (Dim 1, sign $+$, type A)]
\begin{verbatim}
Question: Please tell me whether you think the following action can always be
justified, never be justified, or something in between, using this scale where 1
means "never justifiable" and 10 means "always justifiable."

Action: Abortion

Give a probability for each of the ten scale points, 1 through 10.
\end{verbatim}
\end{promptbox}

\begin{promptbox}[F063 --- importance of God (Dim 1, sign $-$, type A)]
\begin{verbatim}
Question: How important is God in your life? Please use this scale to indicate.
10 means "very important" and 1 means "not at all important."

Give a probability for each of the ten scale points, 1 through 10.
\end{verbatim}
\end{promptbox}

\begin{promptbox}[G006 --- national pride (Dim 1, sign $+$, type A; omitted from C0)]
\begin{verbatim}
Question: How proud are you to be {nationality}?
'1'. Very proud
'2'. Quite proud
'3'. Not very proud
'4'. Not at all proud
\end{verbatim}
\end{promptbox}

\begin{promptbox}[E018 --- future emphasis on respect for authority (Dim 1, sign $+$, type A)]
\begin{verbatim}
Question: If greater respect for authority were to take place in the near future,
would you say it would be a good thing, a bad thing, or don't you mind?
'1'. Good thing
'2'. Don't mind
'3'. Bad thing
\end{verbatim}
\end{promptbox}

\begin{promptbox}[A008 --- feeling of happiness (Dim 2, sign $-$, type A), Wave 4]
\begin{verbatim}
Question: Taking all things together, would you say you are:
'1'. Very happy
'2'. Quite happy
'3'. Not very happy
'4'. Not at all happy
\end{verbatim}
\end{promptbox}
\begin{promptbox}[A008 --- feeling of happiness, Waves 5, 6, and 7]
\begin{verbatim}
Question: Taking all things together, would you say you are:
'1'. Very happy
'2'. Rather happy
'3'. Not very happy
'4'. Not at all happy
\end{verbatim}
\end{promptbox}

\begin{promptbox}[E025 --- signing a petition (Dim 2, sign $-$, type A)]
\begin{verbatim}
Question: Here is a form of political action that people can take. Please tell me
whether you have actually done it, whether you might do it, or would never under
any circumstances do it: signing a petition.
'1'. Have done
'2'. Might do
'3'. Would never do
\end{verbatim}
\end{promptbox}

\begin{promptbox}[A165 --- interpersonal trust (Dim 2, sign $-$, type A)]
\begin{verbatim}
Question: Generally speaking, would you say that most people can be trusted or
that you need to be very careful in dealing with people?
'1'. Most people can be trusted
'2'. Need to be very careful
\end{verbatim}
\end{promptbox}

\begin{promptbox}[CARDSORT --- child qualities (yields A029, A039, A040, A042; Dim 1, type B), Wave 7 (11 cards, cap 5): full item as presented]
\begin{verbatim}
Question: Here is a list of qualities that children can be encouraged to learn
at home. Which, if any, do you consider to be especially important? Please
choose up to five!

1. Good manners
2. Independence
3. Hard work
4. Feeling of responsibility
5. Imagination
6. Tolerance and respect for other people
7. Thrift, saving money and things
8. Determination, perseverance
9. Religious faith
10. Not being selfish (unselfishness)
11. Obedience

This question is an exception to the sum-to-one rule stated above: do NOT make
these values sum to 1. For each quality, give the percentage of the population
who would name it.
Respond with only a JSON object mapping each quality number to a percentage
between 0 and 100. Because respondents may choose at most five, the percentages
should sum to no more than 500.
\end{verbatim}
\end{promptbox}

\begin{promptbox}[CARDSORT, Waves 4 and 5 (10 cards, cap 5): deck only; Wave 6 is this deck with ``Self-expression'' appended at position 11]
\begin{verbatim}
1. Independence
2. Hard work
3. Feeling of responsibility
4. Imagination
5. Tolerance and respect for other people
6. Thrift, saving money and things
7. Determination, perseverance
8. Religious faith
9. Not being selfish (unselfishness)
10. Obedience
\end{verbatim}
\end{promptbox}

\begin{promptbox}[Y002 --- post-materialism, four-item index (Dim 2, sign $+$, type C): full item as presented]
\begin{verbatim}
Question: People sometimes talk about what the aims of this country should be for
the next ten years. If you had to choose, which one of the things on this card
would you say is most important? And which would be the next most important?

1. Maintaining order in the nation
2. Giving people more say in important government decisions
3. Fighting rising prices
4. Protecting freedom of speech

Respond with only a JSON object mapping each ordered pair "first,second" to a
probability, across all twelve possible pairs (e.g. "1,2", "1,3", ... "4,3").
Probabilities must sum to 1. No other text.
\end{verbatim}
\end{promptbox}

\begin{table}[H]
\small
\sbox{\tabbox}{%
\begin{tabular}{lL{3.4cm}L{3.4cm}L{4.6cm}}
\toprule
Condition & Prompt names & Purpose & Cells per model \\
\midrule
C0 & neither country nor year (``a population of adults'') &
country-agnostic prior &
40 (one elicitation, scored under each country's item base) \\
C1 & country only & unprompted present-tense representation & 40 \\
C2 & country and fieldwork year & representation matched to a specific wave
& 134 \\
\bottomrule
\end{tabular}}%
\begin{minipage}{\wd\tabbox}
\caption{Elicitation conditions.}%
\label{tab:conditions}
\usebox{\tabbox}\par
\tabnote{National pride (G006) is undefined without a country and is omitted from C0, which therefore rests on twelve components rather than thirteen. C0 is a single elicitation per model: the 40 rows per model arise because the same stated distributions are scored under each country's indicator base, not because the model was queried 40 times.}
\end{minipage}
\end{table}

\begin{table}[H]
\caption{Condition wrappers, prepended to the item stem.}%
\label{tab:wrappers}
\small
\begin{tabular}{lL{7.6cm}L{3.6cm}}
\toprule
Condition & Wrapper text & Scope \\
\midrule
C0 & Consider a population of adults. They are asked the following survey
question. & once per model; no country, no year \\
C1 & Consider adults living in \{country\}. They are asked the following survey
question. & full grid \\
C2 & Consider adults living in \{country\} in \{year\}. They are asked the
following survey question. & full grid, one call per retained wave year \\
\bottomrule
\end{tabular}
\end{table}

\begin{table}[H]
\caption{Elicitation and scoring completeness.}%
\label{tab:completeness}
\small
\begin{tabular}{L{6.4cm}L{7.4cm}}
\toprule
Quantity & Value \\
\midrule
Elicitation files & 160 (4 models $\times$ 40 countries) \\
Option-level rows & 58,188 \\
Cells per model & 2,100 \\
Independent draws per cell (\texttt{k\_used}) & 10, for every cell in all four models \\
Model coordinate rows & 856 (160 C0 $+$ 160 C1 $+$ 536 C2) \\
Model coordinate range, Dim 1 & [\mn0.52, 0.70] (truth [\mn0.65, 0.83]) \\
Model coordinate range, Dim 2 & [\mn0.52, 0.85] (truth [\mn0.84, 0.75]) \\
\bottomrule
\end{tabular}
\end{table}

\section*{S5. Scoring model outputs onto the cultural map}

\begin{table}[H]
\caption{Model item means against WVS pooled item means.}%
\label{tab:facevalidity}
\small
\begin{tabular}{lcclcc}
\toprule
Item & Model mean & WVS $\mu$ & Item & Model mean & WVS $\mu$ \\
\midrule
A008 & 2.01 & 1.89  & E018 & 1.48 & 1.55 \\
A029 & 0.345 & 0.497 & E025 & 1.94 & 2.20 \\
A039 & 0.321 & 0.365 & F063 & 7.25 & 7.67 \\
A040 & 0.312 & 0.410 & F118 & 4.32 & 3.58 \\
A042 & 0.313 & 0.379 & F120 & 4.14 & 3.23 \\
A165 & 1.70 & 1.72  & G006 & 1.53 & 1.55 \\
Y002 & 1.80 & 1.78  &      &      &      \\
\bottomrule
\end{tabular}
\end{table}

\section*{S6. Snapshot alignment}

\begin{table}[H]
\footnotesize
\setlength{\tabcolsep}{4pt}
\sbox{\tabbox}{%
\begin{tabular}{L{2.6cm}C{1.3cm}C{1.3cm}C{1.3cm}C{1.3cm}C{1.3cm}C{1.3cm}C{1.3cm}C{1.3cm}}
\toprule
Model & C2 mean & C2 median & C2 Dim 1 bias & C2 Dim 2 bias & C1 mean &
C1 median & C1 Dim 1 bias & C1 Dim 2 bias \\
\midrule
Gemini 3.6 Flash & 0.151 & 0.125 & \mn0.067 & \mn0.013 & 0.153 & 0.138 & \mn0.047 & 0.000 \\
Claude Opus 4.8 & 0.182 & 0.161 & \mn0.040 & 0.052 & 0.177 & 0.156 & \mn0.049 & 0.043 \\
GPT-5.5 & 0.184 & 0.166 & \mn0.047 & 0.073 & 0.193 & 0.172 & \mn0.046 & 0.085 \\
Qwen3.6-Plus & 0.231 & 0.194 & 0.058 & 0.087 & 0.238 & 0.200 & 0.067 & 0.118 \\
\bottomrule
\end{tabular}}%
\begin{minipage}{\wd\tabbox}
\caption{Distance to each country's most recent WVS coordinate,
and signed per-axis bias, under both conditions (\textit{n} = 40).}%
\label{tab:c1c2}
\usebox{\tabbox}\par
\tabnote{Distance is Euclidean, in standardized cultural-map units, to the country's most recent WVS coordinate. Positive bias means the model places countries higher on the axis than WVS records, more secular on Dim 1 and more self-expressive on Dim 2.}
\end{minipage}
\end{table}

\begin{table}[H]
\footnotesize
\sbox{\tabbox}{%
\begin{tabular}{lcccc}
\toprule
& \multicolumn{2}{c}{Dim 1 (tradition--secular)}
& \multicolumn{2}{c}{Dim 2 (survival--self-expression)} \\
\cmidrule(lr){2-3}\cmidrule(lr){4-5}
Model & C1 (country) & C2 (country + year)
      & C1 (country) & C2 (country + year) \\
\midrule
Gemini 3.6 Flash & \mn0.047 [\mn0.093, \mn0.004]$^{*}$ & \mn0.067 [\mn0.112, \mn0.027]$^{*}$
                 & 0.000 [\mn0.029, 0.031] & \mn0.013 [\mn0.043, 0.017] \\
Claude Opus 4.8  & \mn0.049 [\mn0.098, \mn0.001]$^{*}$ & \mn0.040 [\mn0.094, 0.012]
                 & 0.043 [0.010, 0.076]$^{*}$ & 0.052 [0.020, 0.085]$^{*}$ \\
GPT-5.5          & \mn0.046 [\mn0.092, \mn0.002]$^{*}$ & \mn0.047 [\mn0.091, \mn0.005]$^{*}$
                 & 0.085 [0.048, 0.124]$^{*}$ & 0.073 [0.036, 0.110]$^{*}$ \\
Qwen3.6-Plus     & 0.067 [0.018, 0.114]$^{*}$ & 0.058 [0.006, 0.107]$^{*}$
                 & 0.118 [0.067, 0.169]$^{*}$ & 0.087 [0.036, 0.136]$^{*}$ \\
\bottomrule
\end{tabular}}%
\begin{minipage}{\wd\tabbox}
\caption{Signed per-axis bias by anchoring condition, with paired-bootstrap
95\% CIs (latest wave, \textit{n} = 40).}%
\label{tab:axisbias}
\usebox{\tabbox}\par
\tabnote{Positive means the model places countries higher on the axis than WVS
records, more secular on Dim 1, more self-expressive on Dim 2. Point estimates are
the mean signed gap across the 40 countries; brackets are 2.5th and 97.5th
percentiles of 2{,}000 country-resampled bootstrap replicates, drawn on the same
resamples across models. $^{*}$ marks an interval excluding zero.}
\end{minipage}
\end{table}

\begin{table}[H]
\small
\sbox{\tabbox}{%
\begin{tabular}{lcccc}
\toprule
& \multicolumn{2}{c}{C1 (country)} & \multicolumn{2}{c}{C2 (country + year)} \\
\cmidrule(lr){2-3}\cmidrule(lr){4-5}
Comparison & Difference & 95\% CI & Difference & 95\% CI \\
\midrule
Qwen \mn{} Claude   & 0.116$^{*}$  & [0.083, 0.146]        & 0.098$^{*}$  & [0.063, 0.132] \\
Qwen \mn{} Gemini   & 0.114$^{*}$  & [0.078, 0.147]        & 0.125$^{*}$  & [0.086, 0.162] \\
Qwen \mn{} GPT      & 0.113$^{*}$  & [0.089, 0.137]        & 0.105$^{*}$  & [0.080, 0.130] \\
Claude \mn{} Gemini & \mn0.002     & [\mn0.053, 0.047]     & 0.027        & [\mn0.025, 0.078] \\
Claude \mn{} GPT    & \mn0.003     & [\mn0.044, 0.039]     & 0.007        & [\mn0.034, 0.049] \\
Gemini \mn{} GPT    & \mn0.001     & [\mn0.018, 0.019]     & \mn0.020$^{*}$ & [\mn0.038, \mn0.001] \\
\bottomrule
\end{tabular}}%
\begin{minipage}{\wd\tabbox}
\caption{Pairwise differences in Dim 1 bias (row \mn{} column), by anchoring
condition, paired bootstrap.}%
\label{tab:pairdim1}
\usebox{\tabbox}\par
\end{minipage}
\end{table}

\begin{table}[H]
\small
\sbox{\tabbox}{%
\begin{tabular}{lcccc}
\toprule
& \multicolumn{2}{c}{C1 (country)} & \multicolumn{2}{c}{C2 (country + year)} \\
\cmidrule(lr){2-3}\cmidrule(lr){4-5}
Comparison & Difference & 95\% CI & Difference & 95\% CI \\
\midrule
Qwen \mn{} Claude   & 0.061$^{*}$    & [0.026, 0.096]       & 0.049$^{*}$    & [0.016, 0.082] \\
Qwen \mn{} Gemini   & 0.085$^{*}$    & [0.040, 0.129]       & 0.080$^{*}$    & [0.033, 0.126] \\
Qwen \mn{} GPT      & 0.045$^{*}$    & [0.012, 0.078]       & 0.047$^{*}$    & [0.015, 0.080] \\
Claude \mn{} Gemini & 0.024          & [\mn0.007, 0.055]    & 0.031          & [\mn0.006, 0.067] \\
Claude \mn{} GPT    & \mn0.016       & [\mn0.039, 0.006]    & \mn0.002       & [\mn0.028, 0.023] \\
Gemini \mn{} GPT    & \mn0.040$^{*}$ & [\mn0.065, \mn0.017] & \mn0.033$^{*}$ & [\mn0.060, \mn0.006] \\
\bottomrule
\end{tabular}}%
\begin{minipage}{\wd\tabbox}
\caption{Pairwise differences in mean placement distance (row \mn{} column), by
anchoring condition.}%
\label{tab:pairdist}
\usebox{\tabbox}\par
\tabnote{Distance is Euclidean, to the country's most recent WVS coordinate, so
a negative difference means the row model is closer and therefore more accurate.}
\end{minipage}
\end{table}

\begin{table}[H]
\small
\sbox{\tabbox}{%
\begin{tabular}{lcccc}
\toprule
& \multicolumn{2}{c}{C1 (country)} & \multicolumn{2}{c}{C2 (country + year)} \\
\cmidrule(lr){2-3}\cmidrule(lr){4-5}
Model & SD & Ratio to truth & SD & Ratio to truth \\
\midrule
\multicolumn{5}{l}{\textit{Dim 1.; truth SD = 0.327}} \\
Gemini 3.6 Flash & 0.358 & 1.094 [0.979, 1.233]        & 0.361 & 1.106 [1.006, 1.225]$^{*}$ \\
Claude Opus 4.8  & 0.213 & 0.653 [0.585, 0.731]$^{*}$  & 0.213 & 0.652 [0.564, 0.752]$^{*}$ \\
GPT-5.5          & 0.335 & 1.024 [0.901, 1.168]        & 0.328 & 1.003 [0.885, 1.145] \\
Qwen3.6-Plus     & 0.289 & 0.887 [0.779, 1.006]        & 0.287 & 0.881 [0.774, 1.001] \\
\addlinespace
\multicolumn{5}{l}{\textit{Dim 2.; truth SD = 0.338}} \\
Gemini 3.6 Flash & 0.327 & 0.966 [0.881, 1.064]        & 0.337 & 0.994 [0.910, 1.095] \\
Claude Opus 4.8  & 0.301 & 0.889 [0.801, 0.999]$^{*}$  & 0.319 & 0.946 [0.856, 1.043] \\
GPT-5.5          & 0.307 & 0.909 [0.811, 1.039]        & 0.331 & 0.981 [0.892, 1.105] \\
Qwen3.6-Plus     & 0.262 & 0.774 [0.666, 0.927]$^{*}$  & 0.257 & 0.760 [0.654, 0.907]$^{*}$ \\
\bottomrule
\end{tabular}}%
\begin{minipage}{\wd\tabbox}
\caption{Cross-country spread of coordinates at each country's latest wave,
model against truth, by anchoring condition (\textit{n} = 40).}%
\label{tab:spread}
\usebox{\tabbox}\par
\tabnote{The ratio is the cross-country SD of a model's coordinates divided by
the SD of the empirical benchmark on the same axis, so 1 means the model
separates countries as widely as the survey record does and below 1 means it
places them closer together. Intervals are 2.5th and 97.5th percentiles of
2{,}000 country-resampled bootstrap replicates, with numerator and denominator
recomputed on each resample.}
\end{minipage}
\end{table}

\begin{table}[H]
\begin{minipage}{0.78\linewidth}
\caption{Model placement before and after naming the country.}
\label{tab:c0baseline}

\footnotesize
\setlength{\tabcolsep}{4pt}

\begin{tabular}{L{2.7cm}C{1.2cm}C{1.2cm}C{3.1cm}C{1.1cm}C{3.1cm}}
\toprule
Model & C0 $\rightarrow$ truth & C1 $\rightarrow$ truth &
Improvement from naming the country & Clears 0 &
C0 $\rightarrow$ C1 displacement \\
\midrule
Gemini 3.6 Flash & 0.532 & 0.153 & 0.379 [0.307, 0.457] & yes & 0.545 [0.469, 0.622] \\
Claude Opus 4.8 & 0.493 & 0.177 & 0.316 [0.256, 0.378] & yes & 0.390 [0.333, 0.445] \\
GPT-5.5 & 0.550 & 0.193 & 0.357 [0.284, 0.432] & yes & 0.486 [0.409, 0.565] \\
Qwen3.6-Plus & 0.616 & 0.238 & 0.378 [0.296, 0.463] & yes & 0.453 [0.372, 0.536] \\
\bottomrule
\end{tabular}

\begin{tablenotes}[flushleft]
\footnotesize
\item \textit{Note.} Values are means across 40 countries; brackets show 95\% bootstrap confidence intervals. Improvement measures how much closer the model is to the surveyed position after the country is named. C0-to-C1 displacement is a distance and is therefore not tested against zero.
\end{tablenotes}

\end{minipage}
\end{table}

\begin{table}[H]
\caption{Countries closest to each model's country-agnostic baseline.}%
\label{tab:c1shift}
\small
\begin{tabular}{lL{10cm}}
\toprule
Model & Five least-differentiated countries (C0 $\rightarrow$ C1 distance) \\
\midrule
Gemini & Uruguay (0.051), United States (0.080), Argentina (0.199), Taiwan (0.202), Chile (0.241) \\
GPT-5.5 & Argentina (0.073), Uruguay (0.074), United States (0.102), Singapore (0.119), Chile (0.133) \\
Qwen & Taiwan (0.054), Singapore (0.056), United States (0.111), Uruguay (0.129), Argentina (0.160) \\
Claude & Argentina (0.073), Uruguay (0.085), Taiwan (0.121), Chile (0.143), Thailand (0.170) \\
\bottomrule
\end{tabular}
\end{table}

\begin{table}[H]
\small
\sbox{\tabbox}{%
\begin{tabular}{clcccccc}
\toprule
Rank & Country & Class & Gemini & Claude & GPT & Qwen & Mean \\
\midrule
1 & Iran & ambiguous & 0.103 & 0.068 & 0.091 & 0.104 & 0.091 \\
2 & Brazil & directional & 0.131 & 0.059 & 0.014 & 0.191 & 0.099 \\
3 & Kyrgyzstan & directional & 0.144 & 0.037 & 0.077 & 0.149 & 0.102 \\
4 & Russia & stable & 0.036 & 0.090 & 0.109 & 0.183 & 0.104 \\
5 & Philippines & stable & 0.187 & 0.045 & 0.127 & 0.076 & 0.109 \\
\ldots & & & & & & & \\
36 & Egypt & ambiguous & 0.119 & 0.303 & 0.307 & 0.479 & 0.302 \\
37 & Taiwan & directional & 0.217 & 0.348 & 0.346 & 0.318 & 0.307 \\
38 & Japan & reversal & 0.255 & 0.466 & 0.283 & 0.347 & 0.338 \\
39 & South Korea & reversal & 0.209 & 0.443 & 0.409 & 0.352 & 0.353 \\
40 & Thailand & directional & 0.462 & 0.348 & 0.410 & 0.391 & 0.403 \\
\bottomrule
\end{tabular}}%
\begin{minipage}{\wd\tabbox}
\caption{Distance to the most recent WVS coordinate, per country
and model, sorted by the mean across models.}%
\label{tab:percountryerror}
\usebox{\tabbox}\par
\tabnote{Full 40-row table deposited as
\texttt{SI\_per\_country\_level\_distance.csv}; the extremes are shown here.}
\end{minipage}
\end{table}

\begin{figure}[htbp]
    \centering
    \includegraphics[width=.75\linewidth]{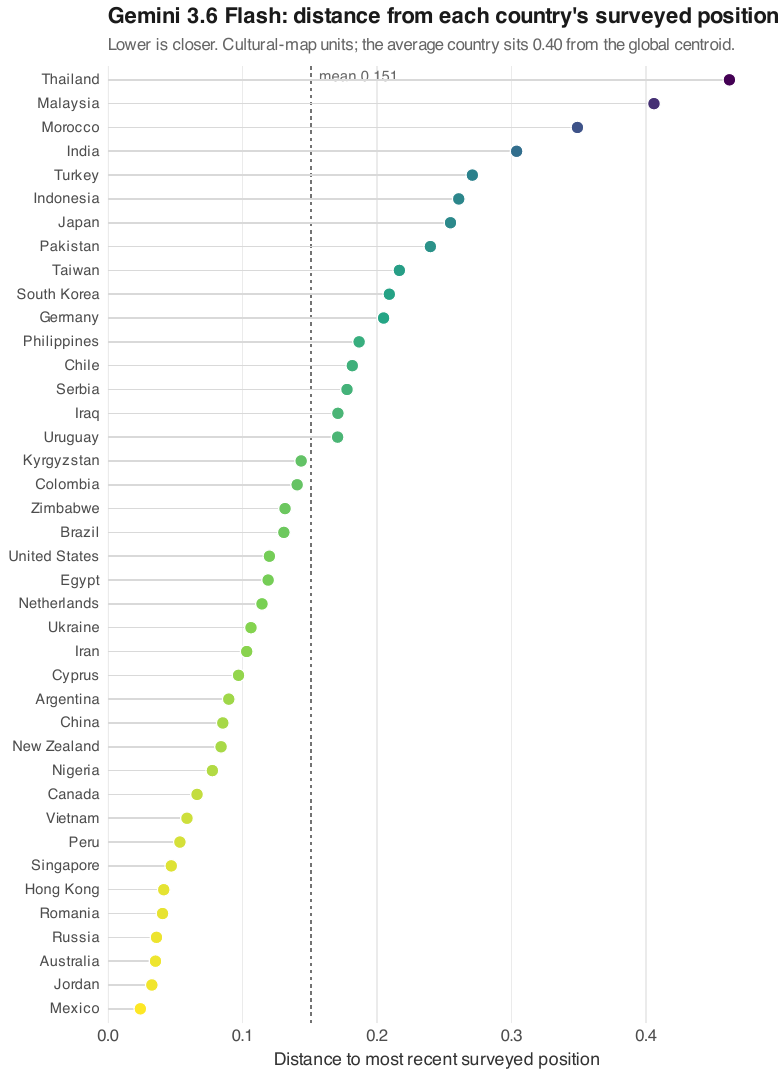}
    \caption{\textbf{Placement error by country, Gemini 3.6 Flash.} Distance
    between the model's placement and each country's surveyed position at its
    most recent wave, in cultural-map units, where the average country sits 0.40
    from the global centroid. Lower is closer. Countries are ordered by this
    model's error, so the ordering differs across models; point shading also
    tracks distance, darker for larger errors. The dashed line marks the model's
    mean of 0.151. All four per-country figures share a common horizontal scale.}
    \label{figS-country-gemini}
\end{figure}

\begin{figure}[htbp]
    \centering
    \includegraphics[width=0.75\linewidth]{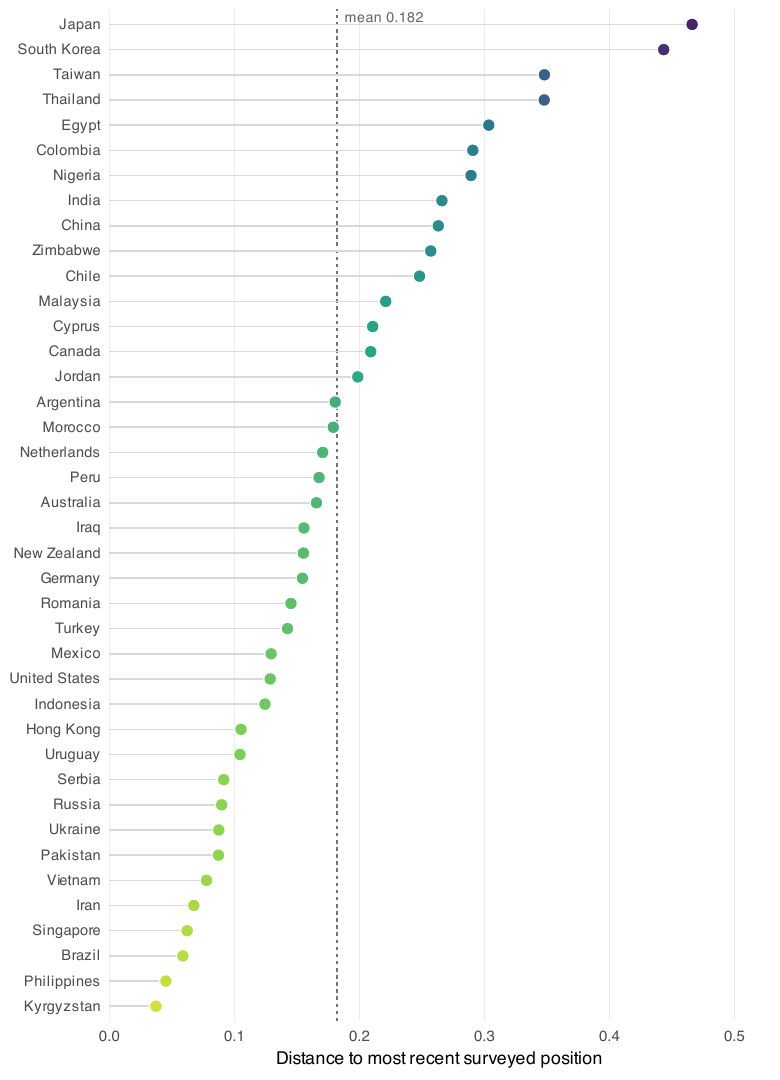}
    \caption{\textbf{Placement error by country, Claude Opus 4.8.} Distance
    between the model's placement and each country's surveyed position at its
    most recent wave, in cultural-map units, where the average country sits 0.40
    from the global centroid. Lower is closer. Countries are ordered by this
    model's error, so the ordering differs across models; point shading also
    tracks distance, darker for larger errors. The dashed line marks the model's
    mean of 0.182. All four per-country figures share a common horizontal scale.}
    \label{figS-country-claude}
\end{figure}

\begin{figure}[htbp]
    \centering
    \includegraphics[width=0.75\linewidth]{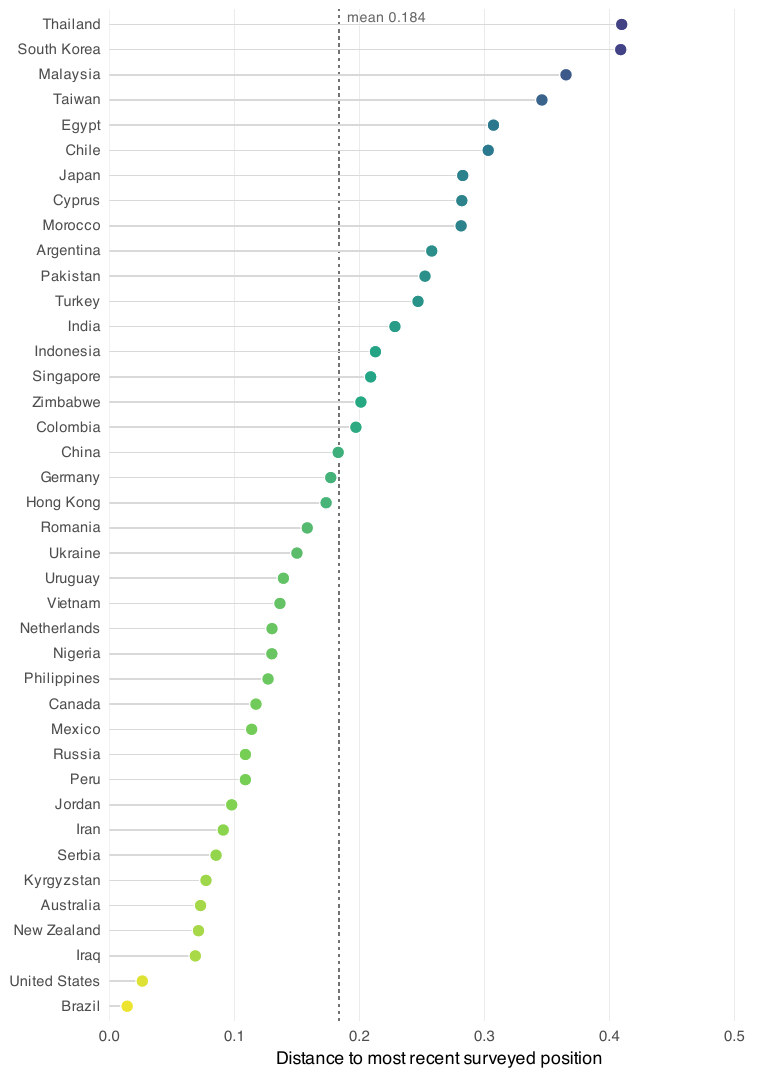}
    \caption{\textbf{Placement error by country, GPT-5.5.} Distance between the
    model's placement and each country's surveyed position at its most recent
    wave, in cultural-map units, where the average country sits 0.40 from the
    global centroid. Lower is closer. Countries are ordered by this model's
    error, so the ordering differs across models; point shading also tracks
    distance, darker for larger errors. The dashed line marks the model's mean of
    0.184. All four per-country figures share a common horizontal scale.}
    \label{figS-country-gpt}
\end{figure}

\begin{figure}[htbp]
    \centering
    \includegraphics[width=0.75\linewidth]{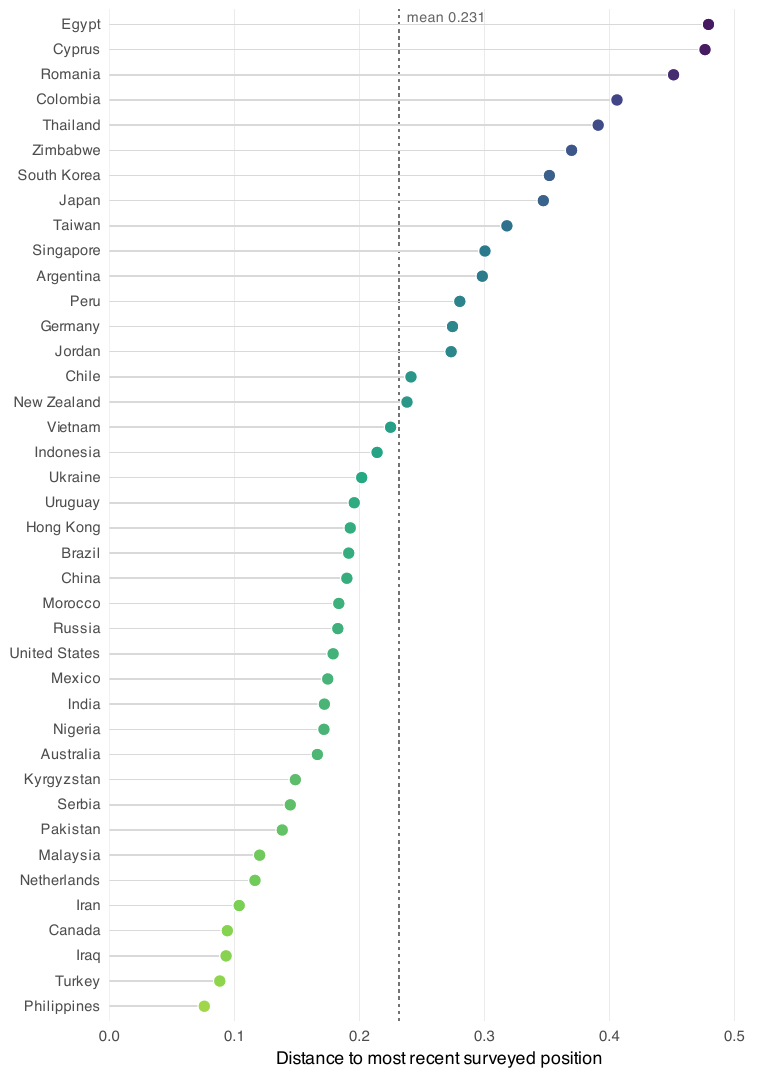}
    \caption{\textbf{Placement error by country, Qwen3.6-Plus.} Distance between
    the model's placement and each country's surveyed position at its most recent
    wave, in cultural-map units, where the average country sits 0.40 from the
    global centroid. Lower is closer. Countries are ordered by this model's
    error, so the ordering differs across models; point shading also tracks
    distance, darker for larger errors. The dashed line marks the model's mean of
    0.231. All four per-country figures share a common horizontal scale.}
    \label{figS-country-qwen}
\end{figure}

\section*{S7. Temporal anchor}

\begin{table}[H]
\small
\sbox{\tabbox}{%
\begin{tabular}{lcccccc}
\toprule
& \multicolumn{4}{c}{Reported set: 36 moving countries}
& \multicolumn{2}{c}{Sensitivity: all 40} \\
\cmidrule(lr){2-5}\cmidrule(lr){6-7}
Model & Mean lag & 95\% CI & Median lag & At present & Mean lag & At present \\
\midrule
Gemini 3.6 Flash & 4.53 & [2.61, 6.58] & 0.0 & 56\% & 4.40 & 58\% \\
Qwen3.6-Plus     & 4.70 & [2.64, 6.97] & 0.0 & 58\% & 4.55 & 60\% \\
GPT-5.5          & 6.00 & [4.00, 8.11] & 5.5 & 42\% & 5.72 & 45\% \\
Claude Opus 4.8  & 7.03 & [4.89, 9.28] & 6.0 & 36\% & 6.65 & 40\% \\
\bottomrule
\end{tabular}}%
\begin{minipage}{\wd\tabbox}
\caption{Implicit temporal anchor: lag behind each country's most recent survey,
under C1.}%
\label{tab:anchor}
\usebox{\tabbox}\par
\tabnote{Lag is measured in years, the age of the wave whose coordinate lies
closest to the model's C1 placement, relative to that country's own most recent
observed wave. A lag of 0 means the model's unprompted placement sits nearest
the most recent survey, and \textit{at present} gives the share of countries for
which this holds. Intervals are 2.5th and 97.5th percentiles of 2{,}000
country-resampled bootstrap replicates; all four exclude zero. The analysis is
restricted to the 36 countries above the movement floor.}
\end{minipage}
\end{table}

\begin{table}[H]
\small
\sbox{\tabbox}{%
\begin{tabular}{lcccc}
\toprule
Model & Wave 4 & Wave 5 & Wave 6 & Wave 7 \\
\midrule
Qwen & 5 & 2 & 8 & 21 \\
Gemini & 3 & 4 & 9 & 20 \\
GPT & 4 & 6 & 11 & 15 \\
Claude & 4 & 8 & 11 & 13 \\
\bottomrule
\end{tabular}}%
\begin{minipage}{\wd\tabbox}
\caption{Number of moving countries whose C1 placement is nearest
each wave (\textit{n} = 36).}%
\label{tab:anchorwave}
\usebox{\tabbox}\par
\tabnote{Nearest wave is the argmin, over that country's observed wave coordinates, of the distance to the model's C1 placement.}
\end{minipage}
\end{table}

\begin{table}[H]
\footnotesize
\setlength{\tabcolsep}{6pt}
\sbox{\tabbox}{%
\begin{tabular}{@{}lccccc@{}}
\toprule
Wave position & Gemini & Claude & GPT & Qwen & \(n\) \\
\midrule
0 (most recent) & \textbf{0.155} & \textbf{0.180} & \textbf{0.194} & \textbf{0.230} & 36 \\
1 & 0.186 & 0.184 & 0.222 & 0.270 & 36 \\
2 & 0.226 & 0.227 & 0.257 & 0.316 & 36 \\
3 (oldest) & 0.256 & 0.295 & 0.266 & 0.335 & 14 \\
\bottomrule
\end{tabular}}%
\begin{minipage}{\wd\tabbox}
\caption{Mean distance from the unprompted (C1) placement to each wave, by wave position.}%
\label{tab:wavepos}
\usebox{\tabbox}\par
\tabnote{Wave position 0 is each country's most recent wave; higher numbers are older waves. Moving countries only. Positions 0--2 are available for all 36; position 3 only for the 14 countries observed at four waves.}
\end{minipage}
\end{table}

\section*{S8. Change alignment}

\begin{table}[H]
\small
\sbox{\tabbox}{%
\begin{tabular}{lcccc}
\toprule
Model & Median magnitude ratio & Median direction cosine &
\% correct direction & \% within 45\dg \\
\midrule
Qwen & 0.902 & 0.772 & 86\% & 56\% \\
Gemini & 0.817 & 0.871 & 83\% & 69\% \\
GPT & 0.672 & 0.708 & 86\% & 50\% \\
Claude & 0.607 & 0.892 & 92\% & 67\% \\
\bottomrule
\end{tabular}}%
\begin{minipage}{\wd\tabbox}
\caption{Start-to-end vector comparison, moving countries
(\textit{n} = 36).}%
\label{tab:netdisp}
\usebox{\tabbox}\par
\tabnote{Moving countries are the 36 above the movement floor. Magnitude ratio is the model's net displacement divided by the country's over the same first-to-last wave span; direction cosine is the cosine of the angle between the two net-displacement vectors. Correct direction is cosine $>$ 0; within 45\dg{} is cosine $>$ 0.71.}
\end{minipage}
\end{table}

\begin{table}[H]
\small
\sbox{\tabbox}{%
\begin{tabular}{lccccc}
\toprule
Decile & 10\% & 25\% & 50\% & 75\% & 90\% \\
\midrule
Ratio & 0.281 & 0.437 & 0.762 & 1.228 & 1.718 \\
\bottomrule
\end{tabular}}%
\begin{minipage}{\wd\tabbox}
\caption{Distribution of the magnitude ratio across all moving
countries and models.}%
\label{tab:magratio}
\usebox{\tabbox}\par
\tabnote{Pooled across the 36 moving countries and all four models. A ratio of 1 means the model travels the true distance.}
\end{minipage}
\end{table}

\begin{table}[H]
\caption{Observed and null mean step cosines, moving countries
(\textit{n} = 36).}%
\label{tab:stepcos}
\footnotesize
\begin{tabular}{lC{1.7cm}C{1.4cm}C{1.7cm}C{1.4cm}C{1.4cm}C{1.7cm}C{2.0cm}}
\toprule
Model & Median observed & Median null & Mean observed & Mean null &
Mean gap & \% beating own null & \% individually significant \\
\midrule
Gemini & 0.454 & 0.244 & 0.404 & 0.281 & 0.123 & 56\% & 0\% \\
Claude & 0.355 & 0.166 & 0.350 & 0.265 & 0.085 & 42\% & 0\% \\
GPT & 0.175 & 0.171 & 0.288 & 0.240 & 0.049 & 33\% & 0\% \\
Qwen & 0.245 & 0.202 & 0.258 & 0.255 & 0.003 & 31\% & 0\% \\
\bottomrule
\end{tabular}
\end{table}

\begin{figure}[htbp]
    \centering
    \includegraphics[width=1\linewidth]{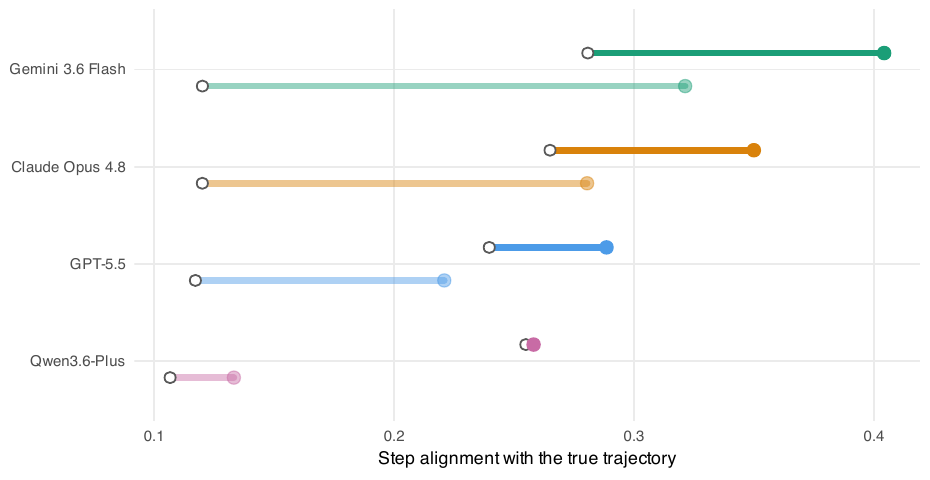}
    \caption{\textbf{Step alignment when a model's steps are matched to the
    right periods versus shuffled onto the wrong ones.} Each bar runs from a
    model's mean alignment under shuffling (open point) to its alignment when
    each step is matched to the period it describes (filled point). A longer bar
    means more of the model's apparent trajectory reflects when change actually
    happened rather than the direction of change alone. Solid bars use all 36
    countries with measured movement; faded bars use the 14 surveyed at four
    waves, where shuffling admits six orderings rather than two and the test has
    more power. Bar length is a point estimate.}
    \label{figS-matched}
\end{figure}

\begin{table}[H]
\small
\sbox{\tabbox}{%
\begin{tabular}{lL{3.4cm}cL{3.4cm}c}
\toprule
Model & All movers (\textit{n} = 36) & Clears 0 &
Four-wave only (\textit{n} = 14) & Clears 0 \\
\midrule
Gemini 3.6 Flash & 0.123 [0.047, 0.203] & yes & 0.204 [0.066, 0.338] & yes \\
Claude Opus 4.8 & 0.085 [0.001, 0.173] & yes & 0.158 [0.017, 0.300] & yes \\
GPT-5.5 & 0.049 [\mn0.019, 0.114] & no & 0.101 [0.012, 0.200] & yes \\
Qwen3.6-Plus & 0.003 [\mn0.051, 0.064] & no & 0.028 [\mn0.060, 0.114] & no \\
\bottomrule
\end{tabular}}%
\begin{minipage}{\wd\tabbox}
\caption{Drift-free gap (observed \mn{} permutation null), full
moving set versus the four-wave subsample where the null is non-degenerate.}%
\label{tab:driftgap}
\usebox{\tabbox}\par
\tabnote{The drift-free gap is the observed mean step cosine minus its permutation-null mean, so it credits only alignment that depends on which period a step is assigned to. Brackets are 95\% CIs from 2{,}000 country-resampled bootstrap replicates.}
\end{minipage}
\end{table}

\begin{table}[H]
\small
\sbox{\tabbox}{%
\begin{tabular}{lL{4.6cm}L{4.6cm}}
\toprule
Comparison & All movers & Four-wave only \\
\midrule
Gemini \mn{} Qwen & 0.121 [0.036, 0.207]$^{*}$ & 0.174 [0.044, 0.331]$^{*}$ \\
Gemini \mn{} GPT & 0.074 [\mn0.014, 0.159] & 0.102 [\mn0.050, 0.260] \\
Gemini \mn{} Claude & 0.039 [\mn0.074, 0.148] & 0.042 [\mn0.137, 0.251] \\
\bottomrule
\end{tabular}}%
\begin{minipage}{\wd\tabbox}
\caption{Between-model differences in the drift-free gap, paired
bootstrap.}%
\label{tab:driftgapdiff}
\usebox{\tabbox}\par
\tabnote{Paired bootstrap over the same country resamples. $^{*}$ marks intervals excluding zero.}
\end{minipage}
\end{table}

\begin{table}[H]
\caption{Drift-free gap under four specifications.}%
\label{tab:leaveout}
\small
\begin{tabular}{lL{2.9cm}L{2.9cm}L{2.9cm}L{2.9cm}}
\toprule
Model & All movers (36) & Four-wave (14) & Excl. reduced base (27) &
Excl. Iraq (35) \\
\midrule
Gemini 3.6 Flash & \textbf{0.123} [0.047, 0.203] \checkmark &
\textbf{0.204} [0.066, 0.338] \checkmark &
\textbf{0.096} [0.008, 0.189] \checkmark &
\textbf{0.128} [0.050, 0.206] \checkmark \\
Claude Opus 4.8 & \textbf{0.085} [0.001, 0.173] \checkmark &
\textbf{0.158} [0.017, 0.300] \checkmark &
0.063 [\mn0.032, 0.155] &
0.076 [\mn0.010, 0.160] \\
GPT-5.5 & 0.049 [\mn0.019, 0.114] &
\textbf{0.101} [0.012, 0.200] \checkmark &
0.029 [\mn0.050, 0.102] &
0.039 [\mn0.030, 0.101] \\
Qwen3.6-Plus & 0.003 [\mn0.051, 0.064] &
0.028 [\mn0.060, 0.114] &
0.020 [\mn0.036, 0.079] &
\mn0.001 [\mn0.057, 0.059] \\
\bottomrule
\end{tabular}
\tabnote{``Reduced base'' excludes the nine countries in Table~\ref{tab:reducedbase}; ``Excl.\ Iraq'' excludes Iraq alone. \checkmark{} marks intervals excluding zero.}
\end{table}

\begin{table}[H]
\caption{Level and rate quantities under the same leave-outs.}%
\label{tab:leaveoutlevel}
\small
\begin{tabular}{lL{3.6cm}L{3.6cm}L{3.6cm}}
\toprule
Quantity & Full (40 / 10) & Excl. reduced base (31 / 9) &
Excl. Iraq (39 / 10) \\
\midrule
Level distance, Gemini & 0.151 [0.120, 0.185] & 0.154 [0.121, 0.195] & 0.150 [0.121, 0.185] \\
Level distance, Claude & 0.182 [0.150, 0.215] & 0.187 [0.151, 0.227] & 0.182 [0.149, 0.215] \\
Level distance, GPT & 0.184 [0.152, 0.215] & 0.183 [0.148, 0.223] & 0.187 [0.158, 0.219] \\
Level distance, Qwen & 0.231 [0.199, 0.266] & 0.236 [0.200, 0.273] & 0.234 [0.202, 0.269] \\
Rate ratio, Gemini & 0.564 [0.425, 0.746] \checkmark & 0.640 [0.364, 0.746] \checkmark & 0.564 [0.425, 0.746] \checkmark \\
Rate ratio, Claude & 0.449 [0.282, 0.615] \checkmark & 0.465 [0.282, 0.643] \checkmark & 0.449 [0.282, 0.615] \checkmark \\
Rate ratio, GPT & 0.576 [0.346, 0.776] \checkmark & 0.621 [0.346, 0.841] \checkmark & 0.576 [0.346, 0.776] \checkmark \\
Rate ratio, Qwen & 0.684 [0.437, 0.792] \checkmark & 0.698 [0.437, 0.836] \checkmark & 0.684 [0.437, 0.792] \checkmark \\
\bottomrule
\end{tabular}
\tabnote{Column headers give the country count for level quantities and for the directional subset used by the rate ratio. \checkmark{} marks intervals excluding 1.}
\end{table}

\begin{table}[H]
\small
\sbox{\tabbox}{%
\begin{tabular}{lcc}
\toprule
Model & Median model movement & Ratio to truth \\
\midrule
Gemini & 0.113 & 1.16 \\
Claude & 0.126 & 1.30 \\
Qwen & 0.145 & 1.49 \\
GPT & 0.170 & 1.75 \\
\bottomrule
\end{tabular}}%
\begin{minipage}{\wd\tabbox}
\caption{Net displacement of the four low-change countries (Russia,
Colombia, Philippines, Cyprus) in each model's representation, against
an observed median net displacement of 0.097.}%
\label{tab:lowchange}
\usebox{\tabbox}\par
\end{minipage}
\end{table}

\begin{table}[H]
\small
\sbox{\tabbox}{%
\begin{tabular}{lL{3.4cm}cL{3.4cm}c}
\toprule
Model & Moving countries (\textit{n} = 36) & Under-moves &
Directional countries (\textit{n} = 10) & Under-moves \\
\midrule
Qwen & 0.902 [0.698, 1.032] & no & 0.684 [0.437, 0.792] & yes \\
Gemini & 0.817 [0.640, 1.178] & no & 0.564 [0.425, 0.746] & yes \\
GPT & 0.672 [0.548, 0.971] & yes & 0.576 [0.346, 0.776] & yes \\
Claude & 0.607 [0.433, 0.855] & yes & 0.449 [0.282, 0.615] & yes \\
\bottomrule
\end{tabular}}%
\begin{minipage}{\wd\tabbox}
\caption{Bootstrap CIs on the median magnitude ratio (1 = model
travels the true distance).}%
\label{tab:rateflat}
\usebox{\tabbox}\par
\tabnote{Magnitude ratio is model net displacement divided by true net displacement. It measures rate cleanly only where the path is close to straight, which is why the ten directional countries give the estimate to report; on reversing or wandering paths the ratio is contaminated by cancellation. ``Under-moves'' means the 95\% interval lies entirely below 1.}
\end{minipage}
\end{table}

\begin{table}[H]
\small
\sbox{\tabbox}{%
\begin{tabular}{lccC{2.6cm}c}
\toprule
Model & Bootstrap median & 95\% CI & Signed-rank median & \textit{p} \\
\midrule
Claude & \mn0.005 & [\mn0.016, 0.006] & \mn0.013 & .314 \\
Gemini & 0.001 & [\mn0.011, 0.014] & \mn0.004 & .675 \\
Qwen & 0.006 & [\mn0.004, 0.017] & 0.014 & .271 \\
GPT & 0.009 & [\mn0.001, 0.019] & 0.016 & \textbf{.026} \\
\bottomrule
\end{tabular}}%
\begin{minipage}{\wd\tabbox}
\caption{Change in distance to truth when the fieldwork year is
named (C1 \mn{} C2; positive = naming the year improves placement).}%
\label{tab:yearnamed}
\usebox{\tabbox}\par
\tabnote{Positive values mean that naming the fieldwork year (C2) brings the placement closer to truth than naming the country alone (C1). Bootstrap over countries, 2{,}000 replicates; the signed-rank test is on the median paired difference. No correction is applied for testing four models.}
\end{minipage}
\end{table}

\section*{S9. Reversal reproduction}

\begin{table}[H]
\caption{All 24 country--model cases across the six verified
reversal countries, ordered by true turn sharpness.}%
\label{tab:revcases}
\footnotesize
\setlength{\tabcolsep}{3.5pt}
\begin{tabular}{llccccccl}
\toprule
Country & Waves & True turn & True mag. & Model & Model turn & Model mag. &
Ratio & Verdict \\
\midrule
India & 4 & 169\dg & 0.37 & Claude & 162\dg & 0.18 & 0.48 & sharp but hollow \\
India & 4 & 169\dg & 0.37 & Gemini & 67\dg & 0.17 & 0.47 & partial bend \\
India & 4 & 169\dg & 0.37 & Qwen & 43\dg & 0.25 & 0.68 & no bend \\
India & 4 & 169\dg & 0.37 & GPT & 15\dg & 0.12 & 0.33 & no bend \\
Jordan & 4 & 169\dg & 0.25 & Gemini & 179\dg & 0.14 & 0.54 & \textbf{REAL} \\
Jordan & 4 & 169\dg & 0.25 & Qwen & 99\dg & 0.06 & 0.24 & partial bend \\
Jordan & 4 & 169\dg & 0.25 & Claude & 72\dg & 0.15 & 0.59 & partial bend \\
Jordan & 4 & 169\dg & 0.25 & GPT & 9\dg & 0.06 & 0.22 & no bend \\
Indonesia & 3 & 165\dg & 0.29 & GPT & 174\dg & 0.16 & 0.54 & \textbf{REAL} \\
Indonesia & 3 & 165\dg & 0.29 & Qwen & 69\dg & 0.15 & 0.50 & partial bend \\
Indonesia & 3 & 165\dg & 0.29 & Gemini & 49\dg & 0.13 & 0.44 & no bend \\
Indonesia & 3 & 165\dg & 0.29 & Claude & 26\dg & 0.16 & 0.54 & no bend \\
Zimbabwe & 3 & 146\dg & 0.49 & Claude & 150\dg & 0.09 & 0.19 & sharp but hollow \\
Zimbabwe & 3 & 146\dg & 0.49 & GPT & 59\dg & 0.08 & 0.15 & no bend \\
Zimbabwe & 3 & 146\dg & 0.49 & Qwen & 36\dg & 0.12 & 0.25 & no bend \\
Zimbabwe & 3 & 146\dg & 0.49 & Gemini & 33\dg & 0.10 & 0.19 & no bend \\
Japan & 4 & 129\dg & 0.24 & Claude & 173\dg & 0.19 & 0.79 & \textbf{REAL} \\
Japan & 4 & 129\dg & 0.24 & Gemini & 169\dg & 0.11 & 0.44 & sharp but hollow \\
Japan & 4 & 129\dg & 0.24 & GPT & 138\dg & 0.15 & 0.61 & \textbf{REAL} \\
Japan & 4 & 129\dg & 0.24 & Qwen & 38\dg & 0.11 & 0.43 & no bend \\
South Korea & 4 & 129\dg & 0.16 & Claude & 105\dg & 0.14 & 0.86 & partial bend \\
South Korea & 4 & 129\dg & 0.16 & Gemini & 47\dg & 0.10 & 0.64 & no bend \\
South Korea & 4 & 129\dg & 0.16 & Qwen & 7\dg & 0.14 & 0.89 & no bend \\
South Korea & 4 & 129\dg & 0.16 & GPT & 2\dg & 0.12 & 0.75 & no bend \\
\bottomrule
\end{tabular}
\end{table}

\begin{table}[H]
\caption{Per-model summary across the six verified reversals.}%
\label{tab:revsummary}
\small
\setlength{\tabcolsep}{4pt}
\begin{tabular}{L{2.6cm}C{1.8cm}C{1.3cm}C{1.3cm}C{1.8cm}C{1.8cm}C{2.0cm}}
\toprule
Model & Sharp turns ($\ge$120\dg) & Genuine & Hollow & Median turn at bend &
Max turn at bend & Median magnitude ratio \\
\midrule
GPT-5.5 & 2 & 2 & 0 & 37\dg & 174\dg & 0.44 \\
Claude Opus 4.8 & 3 & 1 & 2 & 128\dg & 173\dg & 0.56 \\
Gemini 3.6 Flash & 2 & 1 & 1 & 58\dg & 179\dg & 0.45 \\
Qwen3.6-Plus & 0 & 0 & 0 & 40.5\dg & 99\dg & 0.46 \\
\textbf{Truth} & 6 & --- & --- & \textbf{155.5\dg} & 169\dg & 1.00 \\
\bottomrule
\end{tabular}
\end{table}

\begin{figure}
    \centering
    \includegraphics[width=1\linewidth]{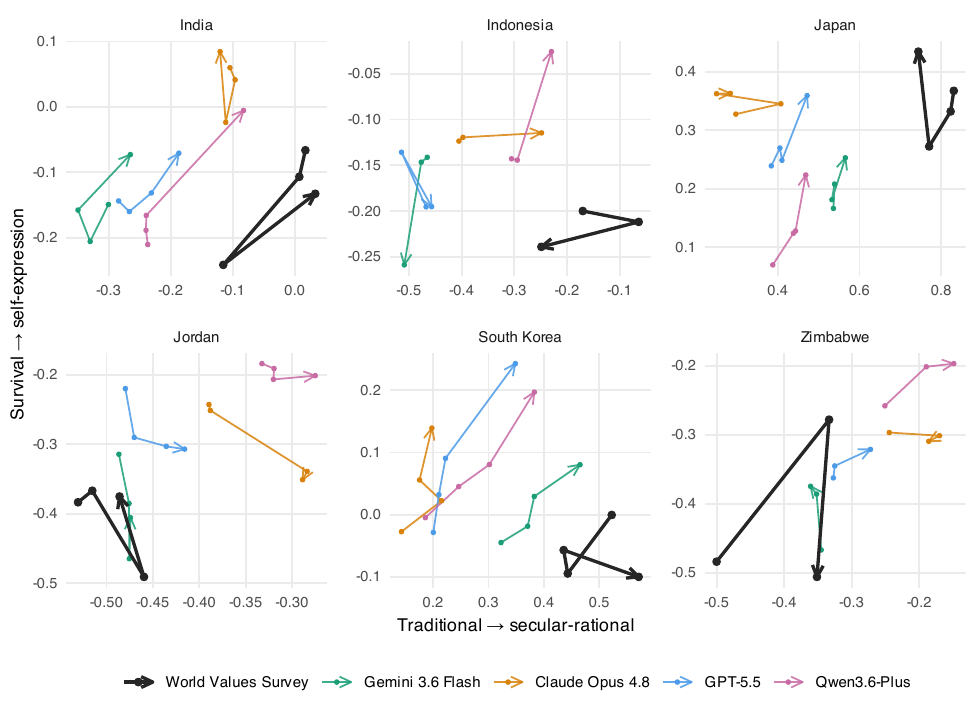}
    \caption{\textbf{Model trajectories through observed cultural reversals.} The six panels correspond to the countries classified as reversals in the WVS data. For each country, the dark path traces its surveyed positions across Waves 4--7, while the lighter paths show the positions produced by each model for the corresponding years. Points indicate survey waves, with arrowheads showing the direction of movement. Because the range of movement differs considerably across countries, axes are scaled separately by panel. Distances should therefore be compared within countries rather than across them.}
    \label{figS-reversal-paths}
\end{figure}

\begin{sidewaystable}[p]
\centering
\captionsetup{width=\textwidth}
\caption{Representation of national cultures across four language models.}%
\label{tab:panel}
\footnotesize
\setlength{\tabcolsep}{6pt}
\renewcommand{\arraystretch}{1.2}
\begin{threeparttable}
\begin{tabular}{@{}l c *{4}{c}@{}}
\toprule
 & Reference & Gemini 3.6 Flash & Claude Opus 4.8 & GPT-5.5 & Qwen3.6-Plus \\
\midrule
\multicolumn{6}{@{}l}{\textbf{Snapshot alignment}\quad\textit{C1 unless noted; 40 countries}} \\
\addlinespace[2pt]
Distance to truth, C1 (country only) & 0 & 0.153 [0.124, 0.182] & 0.177 [0.146, 0.208] & 0.193 [0.160, 0.224] & 0.238 [0.202, 0.278] \\
\quad C2 (country $+$ year) & 0 & 0.151 [0.120, 0.185] & 0.182 [0.150, 0.215] & 0.184 [0.152, 0.215] & 0.231 [0.199, 0.266] \\
\quad C0 (neither named) & 0 & 0.532 & 0.493 & 0.550 & 0.616 \\
Shift when the country is named & --- & 0.545 [0.469, 0.622] & 0.390 [0.333, 0.445] & 0.486 [0.409, 0.565] & 0.453 [0.372, 0.536] \\
\quad of which, gain in accuracy & $>$0 & 0.379 [0.307, 0.457]$^{*}$ & 0.316 [0.256, 0.378]$^{*}$ & 0.357 [0.284, 0.432]$^{*}$ & 0.378 [0.296, 0.463]$^{*}$ \\
Bias, survival--self-expression & 0 & 0.000 [\mn0.029, 0.031] & 0.043 [0.010, 0.076]$^{*}$ & 0.085 [0.048, 0.124]$^{*}$ & 0.118 [0.067, 0.169]$^{*}$ \\
Bias, tradition--secular & 0 & \mn0.047 [\mn0.093, \mn0.004]$^{*}$ & \mn0.049 [\mn0.098, \mn0.001]$^{\dagger}$ & \mn0.046 [\mn0.092, \mn0.002]$^{*}$ & 0.067 [0.018, 0.114]$^{*}$ \\
Spread ratio, tradition--secular & 1 & 1.09 [0.98, 1.23] & 0.65 [0.59, 0.73]$^{*}$ & 1.02 [0.90, 1.17] & 0.89 [0.78, 1.01] \\
Spread ratio, survival--self-expression & 1 & 0.97 [0.88, 1.06] & 0.89 [0.80, 0.99]$^{\dagger}$ & 0.91 [0.81, 1.04] & 0.77 [0.67, 0.93]$^{*}$ \\
\addlinespace[4pt]
\midrule
\multicolumn{6}{@{}l}{\textbf{Implicit temporal anchor}\quad\textit{C1; 36 moving countries}} \\
\addlinespace[2pt]
Lag behind the present, years & 0 & 4.5 [2.6, 6.6]$^{*}$ & 7.0 [4.9, 9.3]$^{*}$ & 6.0 [4.0, 8.1]$^{*}$ & 4.7 [2.6, 7.0]$^{*}$ \\
Lag behind the present, wave-steps & 0 & 0.67 [0.39, 0.94]$^{*}$ & 0.97 [0.69, 1.25]$^{*}$ & 0.89 [0.61, 1.19]$^{*}$ & 0.69 [0.39, 1.03]$^{*}$ \\
Anchored to the most recent wave & 100\% & 56\% & 36\% & 42\% & 58\% \\
\addlinespace[4pt]
\midrule
\multicolumn{6}{@{}l}{\textbf{Change alignment}\quad\textit{C2; moving countries unless noted}} \\
\addlinespace[2pt]
Drift-free timing signal, all movers (36) & $>$0 & 0.123 [0.047, 0.203]$^{*}$ & 0.085 [0.001, 0.173]$^{\dagger}$ & 0.049 [\mn0.019, 0.114] & 0.003 [\mn0.051, 0.064] \\
\quad four-wave countries only (14) & $>$0 & 0.204 [0.066, 0.338]$^{*}$ & 0.158 [0.017, 0.300]$^{*}$ & 0.101 [0.012, 0.200]$^{*}$ & 0.028 [\mn0.060, 0.114] \\
\quad excluding reduced-base countries (27) & $>$0 & 0.096 [0.008, 0.189]$^{*}$ & 0.063 [\mn0.032, 0.155] & 0.029 [\mn0.050, 0.102] & 0.020 [\mn0.036, 0.079] \\
Rate ratio, directional countries (10)\tnote{a} & 1 & 0.56 [0.43, 0.75]$^{*}$ & 0.45 [0.28, 0.62]$^{*}$ & 0.58 [0.35, 0.78]$^{*}$ & 0.68 [0.44, 0.79]$^{*}$ \\
Movement, low-change countries (4)\tnote{a,b} & 0.097 & 0.113 & 0.126 & 0.170 & 0.145 \\
Effect of naming the year (C1 $-$ C2) & 0 & 0.001 [\mn0.011, 0.014] & \mn0.005 [\mn0.016, 0.006] & 0.009 [\mn0.001, 0.019] & 0.006 [\mn0.004, 0.017] \\
\addlinespace[4pt]
\midrule
\multicolumn{6}{@{}l}{\textbf{Reversals}\quad\textit{C2; 6 verified countries}} \\
\addlinespace[2pt]
Sharp turn at the true bend & 6\,/\,6 & 2\,/\,6 & 3\,/\,6 & 2\,/\,6 & 0\,/\,6 \\
Genuine reproduction & 6\,/\,6 & 1\,/\,6 & 1\,/\,6 & 2\,/\,6 & 0\,/\,6 \\
Median turn at the bend & 155.5\dg & 58\dg & 127.5\dg & 37\dg & 40.5\dg \\
\bottomrule
\end{tabular}
\begin{tablenotes}[flushleft]
\footnotesize
\item \textit{Note.} Values are means over countries; brackets are 95\%
confidence intervals from 2,000 bootstrap replicates, resampled identically
across models so that between-model comparisons are paired. \textit{Reference}
is the value a model would show if it represented the quantity perfectly and
$^{*}$ marks an interval excluding it. $^{\dagger}$ marks three intervals that
exclude the reference by less than 0.002 and are better read as marginal than as
decisive; two of the three reverse under the alternative anchoring condition. The C0 row is descriptive and carries no interval;
the accuracy gain below it is the tested quantity. The shift when the country is
named is a Euclidean distance, non-negative by construction, so its interval is
not tested against zero.
Quantities are in standardized cultural-map units unless noted. The axes span
roughly $\pm$0.7 units, and the mean country sits 0.40 units from the global
centroid.
\item[a] Median rather than mean, because the magnitude ratio has a
near-zero denominator on reversing paths and a skewed distribution.
\item[b] Descriptive only; with four countries this row is not bootstrapped and
carries no interval.
\end{tablenotes}
\end{threeparttable}
\end{sidewaystable}

\end{document}